# Topologically engineered 3D printed architectures with superior mechanical strength

**Rushikesh S. Ambekar[1†], Brijesh Kushwaha[1†], Pradeep Sharma[2], Federico Bosia[3], Massimiliano Fraldi[4], Nicola Pugno[5,6*] and Chandra S. Tiwary[1*]**

[1]Metallurgical and Materials Engineering, Indian Institute of Technology Kharagpur, Kharagpur-721302, West Bengal, India

[2]Department of Mechanical Engineering, University of Houston, Houston, TX 77204

[3]Department of Applied Science and Technology, Politecnico di Torino, Italy

[4]Department of Structures for Engineering and Architecture, University of Napoli "Federico II", Italy

[5]Laboratory of Bio-inspired, Bionic, Nano, Meta Materials and Mechanics, Department of Civil, Environmental and Mechanical Engineering, University of Trento, Italy

[6]School of Engineering and Materials Science, Queen Mary University of London, Mile End Road, London, E1 4NS, UK

[†]Equal contribution

## Abstract

Materials that are lightweight yet exhibit superior mechanical properties are of compelling importance for several technological applications that range from aircrafts to household appliances. Lightweight materials allow energy saving and reduce the amount of resources required for manufacturing. Researchers have expended significant effort in the quest for such materials, which require new concepts in both tailoring materials microstructure as well as structural design. Architectured materials, which take advantage of unique structural design, have recently emerged as an exciting avenue to create bespoke combinations of desired macroscopic material responses. In some instances, rather unique structures have emerged from advanced geometrical concepts (e.g. gyroids, menger cubes, or origami/kirigami-based structures), while in others innovation has emerged from mimicking nature in bio-inspired materials (e.g. honeycomb structures, nacre, fish scales etc.). Beyond design, additive manufacturing has enabled the facile fabrication of complex geometrical and bio-inspired architectures, using computer aided design models. The combination of simulations and experiments on these structures have led to an enhancement of mechanical properties, including strength, stiffness and toughness. In this review, we provide a perspective on topologically engineered architectured materials that exhibit optimal mechanical behaviour and can be readily printed using additive manufacturing.

# 1. Introduction

The research community has developed a large class of artificial materials and numerous methods to fabricate them for targeted applications. The selection of appropriate materials is a challenging task, due to the plethora of available possibilities. To facilitate material selection, researchers have prepared databases of material characteristics, including in the form of maps, such as Ashby plots, where critical properties of materials are compared, including metals, polymers, ceramics and composites[1,2]. In this review article, we focus on lightweight and yet strong materials so the key object of our attention is the compressive strength versus density map (**Fig. 1(a)**)[3]. Many conventional strength enhancing techniques are available largely focus on microstructure manipulation, such as cold working, precipitation hardening, solution strengthening and grain boundary strengthening. In precipitation hardening, homogenization of alloys takes place at high temperature, followed by quenching and ageing[4]. The technique of solution strengthening is based on enhancing mechanical properties through dissolution of foreign atoms[5], while grain boundary strengthening exploits changes in average grain size to inhibit the onset of plasticity. Incorporating these solutions at nano scale, such as in the case of nanograin strengthening, can be a valid option, except for costs and the risk of grain growth during heat treatment[6]. When decreasing grain size, the strength of the material increases[7] because dislocations accumulate at grain boundaries[8]. At a critical grain size in the nano range (<10 nm), softening occurs instead of strengthening, due to grain boundary sliding.[9] Ashby plots do not consider an important factor in the design of materials, which is both surface or bulk topology[10]. Researchers have already proven that porous structures have excellent specific strength, additionally to those fabricated using strengthening techniques. In porous structures, the geometry of pores (spherical, cubic, hexagonal, elliptical and octahedron, etc.)[11–13], size of pores (macro, micro and nano)[14–16] and bulk features (curvature, interconnects and directionality)[17–19] all influence mechanical strength. The classification of topology engineered architectures is illustrated in **Fig. 1(b)**.

In this review, we focus on the effect of topology in 3D-printed structures on mechanical properties such as compression strength, tensile strength and energy absorption. Here, we discuss macroscopic-scale structures such as origami and kirigami, and architectures emerging from mathematical models. We also discuss microstructure and porous molecule-inspired architectures, as well as bio-inspired structures.

## 2. Introduction to 3D printing

3D printing techniques are mainly classified in 6 types, i.e. Vat photopolymerization, Material jetting, Powder bed fusion, Material extrusion, Binder jetting and sheet lamination[20]. The Vat Photopolymerization (VP) technique utilizes light in the ultraviolet, or visible light, which selectively polymerizes liquid photosensitive resins. This process has a high accuracy and resolution. Stereolithography (SLA), Digital Light Processing (DLP), Continuous Liquid Interphase Printing (CLIP), and Two-Photon Photopolymerization (2PP) are common printing methods under VP. The printing speed of CLIP and 2PP is 100 to 1000 time higher than that of SLA. The material jetting process involves the use of liquid input materials, and solidification occurs through photopolymerization, cooling, etc. This process is similar to traditional 2D inkjet processes[21]. Polyjet printing, multijet printing and ElectroHydroDynamic (EHD) jetting are examples of material jetting. In EHD, ink is drawn from a nozzle and deposited on the substrate in the form of a thin continuous jet. The printing speed of EHD is of the order of $10^3$–$10^6$ layers per second. In Powder Bed Fusion (PBF), heat sources (laser, infrared radiation etc.) fuse powder particles together to build 3D objects. Selective laser sintering (SLS) is the most widely used process. This process is rather slow due to the point wise laser scanning track, and resolution is limited as compared to VP and MJ due to over-sintering phenomena and finite laser beam diameter. However, the development of new Multi Jet Fusion ( MJF) processes with the combination of PBF and Hewlett-Packard inkjet printing, enable higher speed and higher printing resolution, making the technique cost effective for industrial applications[22]. Material Extrusion involves continuous extrusion of polymer filaments, pellets, etc. through a nozzle. Fused deposition modeling and Direct Ink Writing (DIW) are the most commonly used techniques. FDM is the most affordable and easy technique to use for additive manufacturing processes using thermoplastic filaments as feedstock. DIW can print viscous inks (i.e concentrated polymer solutions and pastes). Advanced DIW systems with multiple nozzles have the capability to print multiple viscous inks[23]. Binder jetting (BJ) is a powder-based inexpensive technique that can fabricate larger parts. Metals and ceramics can also be used in the form of powders. This technique has an advantage over PBF, as it produces large structures without the need for support, but the produced parts are fragile due to the absence of sintering or melting processes. Hence, they require liquid infiltration, high temperature sintering etc. Sheet lamination involves stacking and lamination of thin sheets of material. Laminated object manufacturing (LOM) is the earliest sheet lamination process. Metal sheets, paper, woven fiber composite sheets, ceramic tapes, thermoplastic foils etc. are common printing materials for LOM. This process is not widely used, as it produces significant material waste[24].

## 3. Microscopic scale structures such as origami and kirigami

Origami is an ancient paper folding technology introduced in Japan to create three-dimensional architectures from two-dimensional sheets using suitable folding techniques[25]. Kirigami structures have the same configurations as origami with additional distinctive cutting patterns. These cutting patterns help to achieve particular redistributions of applied loads across the structure[26], and provide ultra-stretchability[27]. Materials whose shape can be programmed as a function of certain environmental conditions such as temperature, pressure, humidity, pH etc. are designated as shape-memory materials[28]. Researchers have merged the two technologies, origami and shape memory materials, to achieve self-foldable systems[29]. Origami facilitate the activation of multi-stable features, and different types of origami structures can be classified according to their stability, either bi-stable or tri-stable. Examples of the most commonly studied bi-stable origami are Yoshimura[30], Waterbomb[31], Miura-ori[32] and square-twist origami structures[33], while tri-stable structures include square-twist origami[34]. Origami structures can be tuned by appropriately folding faces and creases. On the basis of curvature, origami can be classified as zero curvature (Miura-origami tessellation)[35], single curvature (Miura-origami derivatives)[36], double curvature (origami tessellations)[37] and multi curvature structures (Stanford bunny)[38]. Origami structures are popular due to their notable properties like deployment ability, negative Poisson's ratio, bending ability, programmable stiffness, twistable capability, multi-stability and reconfigurability[39,40]. These structures are extensively used in applications like wearable electronics[41], artificial muscles[42], solar arrays in telescopes or space structures[43], autonomous robotics[44], and sensors, antennas and actuators [45]. Traditional designs are static, whereas kirigami/origami exploit structural transformation capabilities through bending, twisting and folding. These transformations can be several orders of magnitude larger compared to traditional designs. Kirigami/origami technology can explore highly complex designs compared to conventional fabrication techniques. Pre-programmed origami designs can successfully replace traditional self-assembly designs with the same functionality[46,47].

## 2.1 Origami-based structures

Kshad *et. al* have fabricated Ron Resch-like origami inspired by polylactic acid via fused deposition modeling for applications such as load and energy dampers. Impact tests showed that Ron Resch-like origami cores provide better energy dissipation than Miura origami cores. Compressive tests revealed that the compressive Young's modulus decreases with expanding folding angle, e.g. the modulus of elasticity for folding angles of 15° (**Fig. 2(a)**), 30° and 60° was 12 MPa, 6.6 MPa and 2.4 MPa. The collapse of Ron-Resch-like structures was observed due to flexural buckling of the interior plate-like faces and then plastic hinges across the planes which were situated in the perpendicular direction to the loading pathway. The shape recovery of the 3-branched RR structure (95%) performed better than the 6-branched RR structure (65%) and the 4-branched RR structure (40%)[48]**.** Manen *et. al* fabricated self-foldable polylactic acid structures via fused deposition modeling for biotechnology and electronics application. To adjust the activation times of the sequential folding, there were two utilized strategies: variation of thickness and porosity. High temperature triggers were utilized to acquire desired pre-programmed 3D shapes of flat structures (**Fig. 2(b)**). As-prepared structures showed higher shrinkage from 13% to 29% at lower (65°C) and higher (95°C) activation temperatures, respectively. PLA-based self-foldable structures showed limitations compared to shape memory materials due to the limited expansion coefficient and non-recoverable deformation[49]. Miura-origami inspired polylactic acid-based shape memory polymer structures were fabricated by Liu *et. al* via fused deposition modeling for applications such as actuators and reconfigurable devices. Compression tests showed that the unfolding condition (97.9 ± 0.3%) of the tessellation structure had a larger shape recovery ratio than the folding condition (95.6 ± 1.1%) (**Fig. 2(c)**). Compression tests also showed that during folding, the tessellation structure (289.6 ± 5.6%) had a larger volume change ratio than the tube structure (228.9 ± 5.7%). DMA tests revealed that temperature significantly affects the recovery force, i.e. as temperature varies from 51°C, to 63°C and to 90°C, the recovery force required for deformation decreases from 2.42 N, to 1.25 N and 0.26 N, respectively[50]. Jo *et. al.* fabricated flexible conductive nanocomposites via fused deposition modeling for energy conversion applications (**Fig. 2(d)**). A mechanical resiliency study showed that the number of layers influences the interconnect strain, e.g. a 4 layer (3500%) origami has larger interconnect strain than 3 layer (2500%) and 2 layer (1600%) origami structures. Experiments show that the Type I origami structure has a greater stretchability at lower force than the Type II origami structure. AgNWs mesh/TPU and perovskite solar modules provide a higher initial areal coverage with 400% reversible system stretchability along with 2500% interconnect stretchability under 100 cycles[51]. Huang *et. al* investigated SiOC ceramic from silicone resin via a direct ink writing technique. A DMA study correlated the rheological properties and the mechanical properties: when the storage modulus ($2 \times 10^5$ Pa) is greater than the loss modulus ($4 \times 10^4$ Pa), the ink solidifies and hence flows with difficulty. Conversely, when the loss modulus is larger than the storage modulus, the slurry can flow easily even at low pressures. A SEM study showed that 400 µm of filament reduces to a diameter of 300 µm after pyrolysis at 1000°C for 2 h[52]. A structure of $ZrO_2$ nanoparticles embedded in a poly (dimethylsiloxane) fabricated via the inkjet printing technique **Fig. 2(e)** was studied by Liu *et. al*. Compression tests revealed that the as-printed ceramic structure had a compression strength up to 547 MPa. It was also observed that the as-printed structure (whose density is 1.6 g/cm$^3$) had a 19 times larger specific compression strength than conventional SiOC foams. Stretchability tests showed that the complex ceramic structure had an elongation of up to 3 times its initial length[43]. Yuan *et. al* investigated origami with complex folding patterns fabricated via multi-material inkjet 3D printing for load bearing applications. The angle between fibers and the loading direction was inversely proportional to the bending angle of the structure; e.g. for 0°, 45° and 67.5° angles between the fiber and the loading direction, the structure displayed 60°, 33° and 6° bending angles, respectively. The fiber arrangement and loading direction was optimized for complex folding patterns (**Fig. 2(f)**) by studying the relation between fiber orientation, folding angle and applied load[53]. Liu *et. al* fabricated origami-based structures via multi-material inkjet 3D printing for consecutive frequency-reconfigurable antennas. Tensile tests showed that rigid polymer-like Verowhite had a larger tensile strength (35 MPa) than soft elastomer-like Tangoblack (0.32 MPa). As-manufactured origami-based antennas utilize an umbrella-like mechanism (**Fig. 2(g)**) for actuation and can function in a frequency range from 0.95 to 1.6 GHz[54]. Origami-inspired shape memory polymer embedded matrix composites for self-assembly were developed by Ge *et. al.* The self-assembly of printed composites was actuated through thermo-mechanical programming, resulting in the conversion of 2D sheets into pre-programed 3D structures. A DMA study revealed that the storage modulus of fibers (~6 MPa – ~1.7 GPa) was greater than the matrix (~0.7 MPa –~900 MPa) in the temperature range from -50°C to 100°C. Similarly, the Young's modulus of the fiber in the fibre-reinforced material (~6 MPa) was larger than the matrix material (~0.7 MPa)[55]. A n-type CNT/Polyvinylpyrrolidone ink as a Terahertz detector for non-destructive testing was reported by Llinas *et. al*. The as-developed origami inspired structure possesses enhanced noise-equivalent power (12 nW/Hz$^{1/2}$) at room temperature[56]. Wang *et. al* investigated classic square-twist origami configurations considering not only material properties but also geometric parameters. Uniaxial tensile tests showed that Transform Mode 1 (0.75 N) had a greater strength than Transform Mode 2 (0.2 N). The printed classic square-twist origami based structure has a unique tri-stable state due to the geometric parameters such as a transformation energy barrier greater than the stored elastic energy in the creases[34]. Zhao *et. al* fabricated origami assemblies via a digital light processing-based 3D printing technique. Compression tests reveal that the thickness of the hinge region has a huge effect on the load bearing capacity, e.g. the latter increases from 2 N, to 4 N, and to 10 N for hinge thicknesses of 400 µm, 600 µm and 900 µm, respectively. Geometric design controls the curvature transformation, load-bearing capacity and spatial expandability[44]. Origami-based PDMS coated polystyrene structures via stereolithography were reported by Deng *et. al*. The self-folding effect of these structures originates from the higher shrinkage of the polystyrene (43% in 5-10 s) between 98°C and 120°C. This shrinkage is responsible for the release of the transformation energy which eventually deforms the coated PDMS layer on the film. The energy absorption efficiency of the as-printed origami structure increases with relative density, e.g. relative densities of 0.159 and 0.054 correspond to energy absorption efficiencies of 2.3 and 8.0 J/cm$^3$ , respectively[57]. Chen *et. al* studied Ron Resch origami pattern-inspired structures for energy absorption applications. Compression tests showed that as-printed structures (**Fig. 3(a)**) had a 6.7 kN compression strength. The Energy absorption efficiency of these structures (8.0 J/cm$^3$) was higher than honeycomb structures (5.4 J/cm$^3$). The length to height ratio of the structures was inversely proportional to the energy absorption efficiency e.g. a *l/h* ratio decrease from 3 to 1 lead to an enhancement in energy absorption from 2.3 J/cm$^3$ to 8.0 J/cm$^3$[58]. Metallic origami structures for electromagnetic absorption were investigated by Cheng *et. al*. The metallic origami structure has a high elastic modulus (55 MPa) and displayed linear behaviour at the first stage of deformation, whereas the second stage displayed plastic behavior. Owing to the efficiency of electromagnetic wave scattering over a broad range of temperatures (20-800°C) the structure exhibited a high reflectivity (<- 10 dB) at 6–7 GHz and 13.8–18 GHz[59].

**2.2 Kirigami based structures**

Nakajima *et. al* studied flexible kirigami-based structures to utilize in wearable devices. Tensile tests showed that the as-printed structure had a larger tensile strength (2.43 MPa) as well as higher elongation at break (183%)[41]. Li *et. al* also studied kirigami-inspired reactive silver ink patters on woven textiles to utilize as an e-textile. Tensile tests showed that kirigami based structures have larger ultimate tensile strain (800%) compared to solid structures (<20%) (**Fig. 3(b)**). The modified textiles showed stable electrical conductivity ($\Delta R/R0 < -20\%$) even at 150% strain[60]. A kirigami-inspired split ring resonator was fabricated by Salim *et. al* using silver nanoparticles as an ink to be utilized as a strain sensor. At 17.24% strain, the resonance frequency increased from 4 to 4.64 GHz. The as-prepared structure had a $4.2 \times 10^7$ Hz/% strain sensitivity and minimum detectable strain level around 0.84%. In kirigami inspired sensors, the relationship between frequency and induced strain provides the sensing capability[61]. Bao *et. al* investigated kirigami-based electrodes for flexible lithium-ion batteries. The flexible electrode was printed using Multi-Walled Carbon NanoTubes embedded in Polyvinylidene ink on a polydimethylsiloxane template. Mechanical tests showed that flexible electrodes had excellent mechanical robustness (500 stretch-release cycles) and a stable discharge capacity (94.5 mA h g$^{-1}$ at 0.3 C after 100 discharge/charge cycles)[62]. Temperature-sensitive kirigami structures for mechanical energy storage were explored by Wang *et. al*. The self-deployment time at 55°C increases with greater length scales. 15-50, 15-60 and 15-70 structures took 0.2 s, 0.5 s and 0.6 s, respectively, to deploy. The folding style was also optimized for quick unfolding e.g. the S-T-1 folding style took a shorter time (2.8 s) than the S-T-4 folding style (1.3 s) (**Fig. 3(c)**). At 55°C, the release of pre-stored air from the folded square-twist structure caused an enhancement in the density of the structure that was responsible for self-sinking behavior[45]. A consolidated schematical summary of literature on origami/kirigami-inspired 3D printed structures is presented in **Table 1**.

## 3. Mathematical model based structures

A Triply Periodic Minimal Surface (TMPS) is an intricate 3D mathematically defined surface with topological homogeneity and zero mean curvature. In 1865, TPMSs were introduced by the German mathematician Schwarz[63]. In 1970, Schoen developed Diamond-based TPMS and Gyroid-based TPMS which are extensively used in current research[64]. There are two techniques to create TPMS-based structures: firstly by thickening the minimal surface to achieve a sheet-based TPMS structure[65], and secondly by filling a solid volume surrounded by a minimal surface to achieve a skeletal-based TPMS structure[66]. These TPMS based structures do not display sharp corners or edges and possess a high level of periodicity in 3D space. They have astonishing properties such as superior mechanical energy absorption, mathematically controllable pore size, heat dissipation, huge surface area, high stiffness to weight ratio and interconnected porosity[67,68]. TPMS based structures are widely utilized in structural engineering[69], tissue engineering[70], blast resistance sandwich structures[71] and sensors for bio-monitoring[72] (**Fig. 4**).

Davoodi *et. al* fabricated TPMS-based Acrylonitrile butadiene styrene moulds via 3D printing for wearable biomonitoring. The as-printed sacrificial mould was used to develop a graphene dip coated porous silicon structure. The silicon sensor showed stable conductivity up to 75% compression strain after 400 deformation cycles (**Fig. 5(a)**)[72]. Similarly, Montazerian *et. al* developed a TPMS-based polylactic acid scaffold for tissue regeneration. The as-printed polylactic acid scaffold was sacrificed to fabricate a porous polydimethylsiloxane scaffold (**Fig. 5(b)**) Compression tests showed that the p-surface scaffold had a greater elastic modulus (0.7 MPa) than the d-surface scaffold (0.3 MPa). At a higher relative density, the p-surface scaffold exhibited greater stiffness than the d-surface scaffold. Scaffolds with radial gradients showed enhanced elastic properties and exhibited higher resistance to deformation before reaching densification[73]. TPMS-based acrylic structures via 3D printing for catalytic substrates were studied by Al-Ketan *et. al*. Compression tests were performed on various TPMS-based structures: Gyroid sheets (**Fig. 5(c)**) displayed a greater compression strength (3.2 MPa) than Gyroids along the loading direction (2.6 MPa) and Gyroid struts (0.15 MPa). Similarly, in the case of primitive structures, struts displayed a greater compression strength (2.5 MPa) than sheets (0.55 MPa). Shear bands were observed before catastrophic failure of the structures. The TPMS-based structure comprised two types of porosity: macroporosity (empty spaces between layers) and microporosity (in the bulk of the substrate)[74]. Al-Ketan *et. al* also investigated TPMS-based microarchitectures with a 100 μm unit cell size. Enhancement in Young's modulus and energy absorption were observed by increasing relative density from 10% to 25%. TPMS sheet-based architectures (**Fig. 5(d)**) exhibited superior strength (30 MPa) and energy absorption compared to TPMS strut-based and octet-truss based architectures (**Fig. 6(a)**)[75]. TPMS-based polymeric cellular structures such as Schoen IWP, Schwarz Primitive, and Neovius structures (**Fig. 5(e)**) were fabricated by Abueidda *et. al*. Mechanical tests showed that IWP-CM and Neovius-CM had higher strength (64 MPa and 98 MPa, respectively) than primitive CM (53 MPa). The deformation of structures and energy absorption (**Fig. 6(b)**) also depends on the properties of the constituent materials: soft materials display buckling or yielding at first, and subsequent densification, whereas hard materials display brittle fracture. For a larger relative density, densification begins at even lower strain in cellular structures[76]. Jia *et. al* studied the effect of shell thickness on TPMS-based Schwarz structures (**Fig. 5(f)**). Compression tests showed that C0T15 structures had higher maximum local von Mises stress (30 MPa) than C0T05 (24 MPa) and C0T10 (25 MPa) structures[77]. *Sajadi et. al* fabricated TPMS-based Schwarz structures (**Fig. 5(g)**) at different length scales. Many types of Schwarz structures such as primitive and gyroid structures were also studied. Impact tests showed that gyroid Schwarz structures have a better capability for energy absorption (14 J g$^{-1}$ cm$^3$) than primitive Schwarz structures (**Fig. 6(c)**). Compression tests showed that primitive Schwarz structures have a greater Young's modulus (9.6 MPa) than gyroid Schwarz structures (7.7 MPa). The gyroid structures with a density of 0.38 g/cm$^3$ had a larger Young's modulus (7.7 MPa) than gyroid structures with a density of 0.58 g/cm$^3$ (4 MPa)[78]. Schwartz diamond graded porous titanium structures were fabricated via laser powder bed fusion for bone implant functionalities. The unit cell size plays a crucial role to achieve better mechanical properties, e.g. an increase in unit cell size from 3.5 mm to 5.5 mm was responsible for a decrease in compression yield strength from 11.43 MPa to 7.73 MPa. A similar trend is also observed in the case of the energy absorption capability, which decreases from 6.06 MJ/mm$^3$ to 4.32 MJ/mm$^3$, respectively[79]. Furthermore, to mimic the localized change in stiffness of bone, researchers have developed Functionally Graded Porous Scaffolds (FGPS) using a selective laser melting technique. Mechanical properties are directly proportional to the strut size of the FGPS, e.g. a strut size variation between 483 and 905 μm can easily tailor the yield strength between 3.79 and 17.75 MPa, and the elastic modulus between 0.28 and 0.59 GPa, respectively[80]. In general, in the post-elastic range, materials can show either brittle or plastic behaviour, characterised in the stress-strain plane by hardening or softening phases and different ductility levels. Al-Ketan *et. al* have investigated unique Gyroid based core-shell structures (**Fig. 5(h)**) where the core is a soft material and the shell is a hard material. Since the difference in elastic modulus between the soft and hard polymer materials is more than 200 MPa, the hard material bears the load and the soft material prevents or delays the crack propagation and hence avoids catastrophic failure. A core-shell structure (0.85 MJ/m$^3$) absorbs more energy than a soft polymer structure (0.10 MJ/m$^3$) and a hard polymer structure (0.60 MJ/m$^3$). The increase in relative density improves the toughness of the structures (**Fig. 6(d)**)[81]. Gyroid based cellular structures (**Fig. 5(i)**) for structural architectures were studied by Abueidda *et. al*. As-printed structures were fabricated without joints or discontinuities, which helps to avoid stress concentration effects. Compression studies showed that the yield strength of the gyroid structure increases with an increase in relative density from 14% (2 MPa) to 46% (14 MPa). Similarly, relative density also has a huge effect on energy absorption at 25% strain e.g. a structure with relative density equal to 0.46 has a higher energy absorption (3000 kJ/m$^3$) than a structure with relative density equal to 0.14 (500 kJ/m$^3$)[69].

Hensleigh *et. al* fabricated octet-truss, gyroid, cubo-octahedron, and Kelvin based micro-architectures via light-based 3D printing. Compression tests showed that bend-dominated gyroid micro-architectures have a larger elastic modulus (2.67 MPa) than stretch-dominated micro-architectures (1.54 MPa). Features down to 10 µm were fabricated along with 60 nm pore sizes[82]. Octet-truss based Metamaterial architectures (**Fig. 5(j)**) fabricated via stereolithography were investigated by Mohsenizadeh *et. al*. A strain rate variation from 4 mm/min to 300 mm/min was shown to increase the compression yield strength of the architecture from 0.05 MPa to 0.24 MPa. The energy absorption efficiency of the octet-truss based architecture was greater (45%) than that of expanded polystyrene (EPS) foams (37%)[83]. Amirkhani *et. al* studied cubic and hexagonal-based structures to design tissue engineering scaffolds. Compression studies showed that a cubic-iso structure (**Fig. 5(k)**) had a higher yield strength (12 MPa) than a cubic structure (8 MPa) and similarly, a hex-out structure had a higher yield strength (12 MPa) than a hex-in structure (9 MPa). The mechanical properties of the structures depend on cell geometry, properties of the material and relative density[70]. The distinctive "boxception" structures (**Fig. 5(l)**) for superior energy absorption were explored by Sajadi *et. al*. Impact tests showed that geometry plays a crucial role in determining the mechanical properties. For example, a boxception structure has a higher energy absorption capability (11.7 J) than a conventional solid structure (10.4 J) (**Fig. 6(e)**). After removal of the uniaxial compression load, the boxception structure has a better recovery to its original state (95%) than the solid structure (65%). The many curved features of the complex structure facilitate the distribution of forces into various smaller components, so that force localization is minimized and hence large-scale damage to the structure prevented[71]. A schematical summary of the consolidated literature on mathematical model-inspired 3D printed structures is presented in **Table 2**.

## 4. Microstructured and porous molecule-inspired structures

Crystal-inspired structures to obtain damage-tolerant architectures were fabricated by Pham *et. al*. Compression tests showed that an eight meta grain-inspired structure has a smaller yield strength (5.1 MPa) than a single grain-inspired structure (5.4 MPa) (**Fig. 7(a)**). Due to the high periodicity of the single grain-inspired structure, catastrophic failure was observed, whereas in the case of the eight meta grain-inspired structure, brittle failure was prevented. The 25 meta-precipitate inspired structure was found to have a 5.3 MPa yield strength, and repeatability of mechanical behavior was also verified. Architectures with conventional microstructures experience catastrophic failure due to the fact that unit cells are oriented in the same direction, but addition of metallurgical features such as precipitates, grain boundaries and different phases lead to an improvement of mechanical properties[84]. Sajadi *et. al* explored tubulane-based high impact resistance structures. The tubulane-based structure was inspired by cross-linked carbon nanotubes, leading to 3 different structures: 8-tetra-(2,2) (structure I), 12-hexa-(3,3) (structure II) and 36-hexa-(3,3) (structure III).

Compression tests showed that structure I had a larger energy absorption capability (35 J/g) than structure II (20 J/g) and structure III (25 J/g). A CT-scan test (**Fig. 7(b)**) shows that structure II (5.6 x $10^4$ mm$^2$) experienced a greater damage area than structure I (1.5 x $10^4$ mm$^2$) and structure III (2.2 x $10^4$ mm$^2$). A solid structure (9.2 x $10^4$ mm$^2$) also experienced a large damage area. The tubulane geometry and porosity are responsible for localized damage, whereas in the solid cube, the crack propagates through the whole structure[85]. An agglomerate-inspired 3D printed structure was reported by Ge *et. al*. Compression tests revealed that spherical agglomerate-inspired structures show higher yield strength (340 N) than cubic tetrahedral-inspired structures (130 N). Sudden fracture was observed in spherical agglomerate-inspired structures at 13% strain, whereas in the case of cubic tetrahedral-inspired structures, two breakage points were observed at 12% strain and 30% strain due to the geometry of the tetrahedral bond[86]. Investigating the mechanical properties of zeolite-templated carbon nano tube networks Ambekar *et. al* reported high compressive strength of 3D printed structures without structural failure. Experimental results showed good agreement with MD simulations. These structures also showed anisotropic behaviour with better Young's Modulus and yield strength in the X-direction than the Z-direction[87]. The array of patterned spherical shells specifically mimic crystal structures (BCC) with negative Poisson's ratio, called auxetic metamaterials. A carbon-nanotube reinforced PA12 metamaterial nanocomposite was developed via selective laser sintering with excellent energy absorption capability (20.42 J g$^{-1}$). BCC-inspired structures with 6-hole and 12-hole unit cells were considered to design an auxetic lattice with relative densities of 0.09 and 0.3, respectively[88]. A schematic summary of the consolidated literature relative to microstructure- and small molecule-inspired 3D printed structures is presented in **Table 2**.

## 5. Bio-inspired structures

Bioinspired design and biomimicry is a new and important field of research[89]. Natural materials often combine exceptional mechanical properties, like strength, stiffness and toughness with low density and widely available constitutive materials, but also display multifunctionality, e.g. smart adhesion, self-healing, self-cleaning, water repellence, etc. Defining aspects of natural materials are structural hierarchy and material heterogeneity, but a great variety of architectures is observed[90]. In recent years, demand for high performance and light materials has increased in many sectors, including the defense, aerospace, automotive, and energy industries[91].

In this field, H.H Sang *et. al* applied this approach to the mimicry of muscles, reporting the system as a good energy absorber[92]. Detailed studies have been performed on honeycomb with a hierarchical structure. This has been found to be far better than a regular honeycomb[91]. For example, tailored honeycomb has good energy absorption properties[93]. Nacre-like structures[94] and bone inspired structures[95] have also shown promising properties in terms of yield strength, energy absorption, Young's modulus etc. Thus, bio inspiration is a good approach to study the design of biological systems and to draw inspiration from them to solve real world problems, if necessary using artificial materials with superior properties, such as graphene or carbon nanotubes.

## 5.1 Bio-inspired cellular structures

L. Sang *et. al* performed dynamic mechanical analysis, mechanical tests and rheology tests to observe the thermal, mechanical and viscoelastic characteristics of 3D printed PLA-PCL/KBF specimens. During compressive loading, all ratios of PLA-PCL/KBF circular honeycombs show elastic deformations with enhanced energy absorption. The incorporation of PCL enhances elasticity of honeycombs and interfacial bonding. At a higher relative density, structures have enhanced strength and energy absorption. Both honeycomb structures (hexagonal & re-entrant) exhibit excellent energy absorption and load-bearing capabilities **Fig. 8(a)**[96]. Longhai Li *et. al* studied glass sponges for the purpose of bio inspiration. The mechanical properties of a glass sponge-inspired tubular structure, named structure I, were compared to those of a honeycomb tube chosen as reference structure. The results show that the lightweight numbers (LWN = Maximum load / weight) of novel bio-inspired structures are greater than those of the honeycomb structures with lower weight of compression (LWN-C), due to reduced mass. The lightweight number of compression (LWN-C) of the bio-inspired structure I is 47.6% greater than that of the honeycomb structure. Simulation of compression tests of structure I and the honeycomb structure show that in the former, stress is maximum at the longitudinal supporting rib which results in local burst damage. On the contrary, in structure I the equivalent stress is mainly located in the 90º and 45º supporting ribs **Fig. 8(b)**[97]. A study by J. Xu *et. al* shows that the compressive behaviour of the structure can be tuned by exploiting micro-topologies. **Fig. 8(c) (i)-(iv)** shows Sq_symtube, Sq_udtube, Kag_udtube and Tri_udtube, structures. 'Sq' stands for square shaped honeycombs; 'udtube' for unidirectionally aligned hollow tubes in each strip; Kag for Kagome shaped honeycomb; Tri for triangular shaped honeycomb. Honeytubes shown greater energy absorption capabilities over honeycombs and balanced tube patterns can ensure accelerated compressive performance of honeytubes. Due to the severe sliding behaviour during compression, specific energy absorption (SEA) values of Sq_udtubes are the lowest among all structures. Instead, for tri_udtube, SEA values are significantly greater than the corresponding honeycombs[98]. A study by S. Kumar *et. al* shows the enhanced energy absorption capability of geometrically tuned honeycombs (re-entrant, irregular hexagonal, and chiral). Energy absorption efficiency values of all three cell topologies for different values of the gradation parameter *α* (defined as the smallest thickness of the wall /highest thickness of wall) indicate that the re-entrant geometry has the highest efficiency of 64.7% for α = 0.35. **Fig. 8(d)** shows the variation of specific energy absorption with *α*. Analysis of architected honeycombs reveals that an increased energy absorption efficiency (up to 90%) can be attained for an optimal relative density (0.33)[93]. S. Yin *et. al,* inspired by hierarchical biological structures, developed an octet, octahedron tetrakaidekahedron **Fig. 8(e)**. General compressive responses of theses specimen show the collapsing of sequential layers in the plastic collapse region. Euler buckling of $1^{st}$ smaller struts (EB1) was the common failure mode for all these three structures[99]. Y. Chen *et. al* designed a new class of hierarchical honeycombs. Their energy absorption and stiffness are 7.5 times and 6.6 times those of regular honeycombs due to introduction of an internal triangular lattice structure. In hierarchical honeycomb structures, local buckling of the triangular lattice increases as relative density increases (**Fig. 8(f)**). These structures display a progressive failure mode during uniaxial compression as well as during cyclic loading[91]. Y. Nian *et. al* studied the energy absorption of novel bio-inspired graded honeycomb-filled circular tubes (BGHCT). Grading affects the bending behaviour and energy absorption. An A-RHT (ascending radial graded honeycomb filled tube) is found to have promising impact energy absorption (per unit mass) at the "knee point (Pareto optimal point)". The energy absorption characteristics of bio-inspired graded honeycomb fillers is also reported. The graded direction plays an important role in the energy absorption capacity and bending behaviour of honeycomb-filled thin-wall structures during bending. The A-RHT has the capability to absorb a greater impact energy (per unit mass) at the knee point compared to the D-RHT (Descending radial honeycomb filled tube) and UHT (Uniform honeycomb tube). Specifically, in the case of an axial graded pattern, tensile failure can considerably reduce energy absorption[100]. From Ashby's chart, it can be concluded that though the mechanical properties of honeytubes are not optimal as compared to other honeycombs, they are better than other lightweight cellular materials.

Tailoring the microstructure design with the aid of AI can fill the space which is currently vacant. Different architected hollow honeytube structures display better specific strength than honeytubes. Hence, they can be potentially used to design and manufacture lightweight materials (**Fig. 9**)[98,101].

**5.2 Bio-inspired composite structures**

Y. Kim *et. al* proposed an analytical model for the prediction of stress distributions within staggered platelet structures. A diagram of fracture patterns displays three distinct propagation zones: soft tip (ST) zones, soft shear (SS) zones and hard platelet (HP) zones. It was found that for high volume fractions of the platelet material, crack propagation patterns are straight, while for low volume fractions, the crack propagation pattern is a zigzag shape **Fig. 10(a)**[102]. are delocalized in nacre-like designs (**Fig. 10(b)**)[103]. Mimicking nacre's multilayer composite structure, P. Tran *et. al* designed three-dimensional Voronoi based composite structures. Results of impulsive loading show that cracks initiate from the edges of the layer and propagate toward central region. The cohesive failures of the top layer are near the end regions, while for the middle layer debonding occurs at the centre of the laminate, and failures are less severe near edges compared to the front layer. The back layer shows no cohesive debonding. Cohesive and adhesive bonds help in minimizing the plastic damage in the composite by absorbing the energy from the imparted shockwave (**Fig. 10(c)**)[104].

G. X. Gu *et. al* developed a nacre-like composite. Velocity vs displacement data match simulations and this nacre-like design prevents the perforation of bullets after impact. Stress and deformation L. S. Dimas *et. al*, inspired by composite natural materials, studied the fracture of bone-like, bio-calcite-like and rotated bone-like structures. In experiments, bio calcite displayed better strength than other structures. Snapshots of fracture propagation show that cracks propagate perpendicularly to the original crack orientation and along the diagonal through the crack tip for bone-like and rotated bone-like structures, respectively. Instead, for a bio calcite-like geometry, fracture propagates with consecutive crack arrest and propagation phases because there is no continuous soft phase for continuous crack advance (**Fig. 10(d)**)[95]. K. Ko *et. al* designed nacre-like laminated composites, employing a Voronoi diagram, with different topologies. Specimen 1 has high maximum flexural load but low deformation, while specimen 3 and 4 have smaller maximum flexural load and larger flexural load. Specimen 2-1 and 2-2 show different maximum loads because of the different V shaped configuration. Specimen 2 is not appropriate for applications because its ductility varies with the location of applied load. Specimen 5 has the best energy absorption capacity (**Fig. 10(e)**)[94]. F. Liu *et. al* studied nacre-like interlocked composites. Single edge notch bending (SENB) fracture tests revealed that two fracture modes are responsible for fracture: interface failure and tablet break. These two fracture modes depend upon the tablet waviness angle, tablet aspect ratio and volume fraction of stiff phase. Results showed that strength, stiffness and toughness increased by 55%, 143%, and 176%, respectively, varying these parameters (**Fig. 10(f)**)[105].

Y. Kim *et. al* also designed isotropic 2D structural composites, indicated as soTstB (Soft tile-stiff boundary) and stTsoB (stiff tile-soft boundary), organized into different topologies: square, hexagonal, circular and quasicrystal. SoTStB performed very well as compared to StTSoB in terms of strength and toughness because it reduced stress concentrations near the crack tip[106]. Inspired by soft and stiff phases of bone and spider silk, M. Lei *et. al* designed a class of periodic 2D elastomer composites with repeat units (**Fig. 11(a).(i)-(ii)**). A stiff glassy polymer provided a honeycomb-like mesh and the elastomer was used to fill the inclusions. Crack propagation is stable in compliant regions and unstable in the stiff regions (**Fig. 11(a).(iii)**). Geometry plays an important role to achieve a wide range of storage modulus values (**Fig. 11(a).(iv)**)[107]. In biological micro-structured materials, fibres play a significant role. Inspired by biological materials, L. Ren *et. al* fabricated fibre-reinforced composites with different fibre arrangements, using doctor blading processes (**Fig. 11(b).(i)-(iv)**). Bio-inspired zigzag and sinusoidal architected designs have shown the largest values of impact energy absorption (**Fig. 11(b).(v)**). Also smaller aspect ratio of fibres have shown lower impact energy[108]. L. Ren *et. al* developed fibre-reinforced composites with a complex internal structure. The suggested approach was successfully utilized to fabricate bio-inspired composites with the help of ''bouligand'' and ''herringbone'' architectures. The influence of the architecture as well as of geometrical parameters on its compression strength and impact toughness were investigated. The results showed that herringbone structure composites with small angles display outstanding compressive resistance. Similarly, ''bouligand'' structure composites display the best impact resistance values (**Fig. 11(c)**)[109]

## 5.3 Other Bio-inspired structures

Chirality refers to the property of non-superposability to the mirror image. Bowen Zheng *et. al* studied the mechanical behaviour of DNA-inspired helical structures. The movement of red markers (**Fig. 12(a).(ii)**) during quasi-static loading shows that 7-(1,2,3,4,5,6,7) and 3-(2,4,6) rotate approximately by the same angle, in the same direction and with no dilation. (Numbers in brackets show the connection between two helixes. (**Fig. 12(a).(i)**) However, 0 - (–) rotates by a relatively small angle in the opposite direction and dilates laterally. So, connections are decisive in determining the deformation mode. The rotation-displacement relation of 7-(1,2,3,4,5,6,7) and 3-(2,4,6) are similar. Other authors also discussed the influence geometrical parameters on the stiffness and deformation modes[110]. C.S. Tiwary *et. al* examined the evolution of complex natural structures like seashells. Their studies discuss two different natural shapes (indicated as shell – 1 and shell – 2). The first is with a diametrically converging localization of stresses and the second with a helicoidally concentric localization of stresses. Also presented is a mechanics based model to explain their evolution (**Fig. 12(b)**)[89]. J. Yang *et. al* presented a comprehensive study of Bi-Directionally Corrugated Panel structures, drawing inspiration from the mantis shrimp telson, with different wave numbers ($N$). Results of compression tests show that with the increment of $N$, the load bearing capability of the as-printed structure improves. Deformation behaviour changes from ductile to brittle as $N$ increases. For $N=4$, small dimples form, for $N=5$ cleavage occurs and for $N=6$ the river pattern is observed on the fracture surface (**Fig. 12(c)**)[111]. H.H. Tsang *et. al* performed an experimental investigation of muscle-inspired hierarchical structures (**Fig. 12(d).(i)-(iii)**), finding that muscles act as a cushioning layer to protect brittle bones. It was reported that their force-displacement behaviour displays nonlinearity and hysteresis. The enhancement of mechanical properties increases with hierarchy. For example, a strength enhancement of 258% and an energy absorption increase of 172% are obtained for the third-order hierarchical structure. These improved properties are observed because of the spatial and temporal delocalization of stress and strain between various levels of hierarchy[92]. M. Schaffner *et. al* developed a 3D printing platform for digital fabrication of silicone-based soft actuators, inspired by plant systems and a muscular hydrostat. The actuators are made up of an elastomeric body whose surface is modified with reinforcing stripes at a well-defined lead angle. Controlling the lead angle of the stiff phase; it is possible to achieve contractile, expanding and twisting motions in the plant system and muscular hydrostat. (**Fig. 12(e).(i)-(iv)**)[112]

DNA is only one of the examples exhibiting helical structures and mechanical chirality. Apart from chirality as crystallographic category, helical structures can be in fact encountered across different scales in a great variety of both natural and artificial (man-made) materials, some of which illustrated in **Fig. 13**. At the microscopic level, the cytoskeleton – the bearing structure of the eukaryotic cell – is for instance constituted by a self-equilibrated assembly of compressed microtubules and a network of tensed actin and intermediate protein filaments organized to store elastic energy to form a so-called a "tensegrity" (a term resulting from haplology of the words "tensional integrity")[113,114]. In such pre-stressed structures, polymerization/depolymerization phenomena allow the microtubules – themselves exhibiting a helical arrangement – to exert forces, thus governing the mechanobiology of the cell, regulating adhesion, migration and duplication through dynamic instability produced by the coexistence of assembly and disassembly at the ends of each single microtubule. The tensegrity principle is itself related to an equilibrium configuration that is often realized by means of geometrically chiral architectures, as one can still observe at any scale, e.g. in helical structures and dynamics of viruses as key mechanism for their activity[115], in plants[116] and bony horns[117] as a consequence of competition between stress and growth that leads to optimized structures. At macroscopic level, both soft and hard biological tissues also present chiral structures, like arterial walls[118,119] and osteons in bone[90,120], where intima, media and adventitia layers of the blood vessels and helically wound lamellae are at the basis of the optimal mechanical performance in term of elasticity and toughness of these biomaterials. Many engineering applications also exploit chiral mechanical structures. In structural engineering, cables of suspended bridges that require extreme axial stiffness and strength but low bending rigidity are always built up as helically wound wires[121]. Hierarchical structures with this same topology have however also been implemented in the fields of materials and mechanical engineering, and examples of the influence of chirality in mechanics have very recently been shown. One example is the case of the analytical[122] hybrid probabilistic-deterministic mechanical strategy used to predict post-elastic response of hierarchically organized strands made of chiral bundles of wires capable to homogeneously redistribute the load when a wire breaks, in which the chirality plays a crucial role in the optimization of the needed overall mechanical performance[123]. Another example is that of elastomeric composites where the reinforcement is composed by cords with helically wound fibers, which are responsible for fatigue life and delamination phenomena occurring at cord-rubber as well as at ply interfaces in tire applications, where it has been definitively demonstrated that the previously neglected effect of chiral arrangement is instead central for the overall response of the material and even allows to envisage new generations of composites[124]. Finally, the vast majority of auxetic foams and metamaterials, which are attracting a growing interest in the scientific community in recent years, are also based on chiral microstructural morphologies[125,126].

R. Martini *et. al* performed a study of bio-inspired protective scales, analyzing arrays of scales with progressively more complex geometries and arrangements, from simple squares with no overlap to complex ganoid scales with overlaps and interlocking features. The corresponding Ashby plot shows that the puncture resistance increases by a factor of 16 and the compliance decreases by factor of 20 (**Fig. 14(a)**). Small changes in geometry are responsible for considerable variations in compliance and puncture resistance[127]. E. Lin *et. al* studied the dependency of the strength, toughness and stiffness of suture interfaces, varying the geometric parameters (**Fig. 14(b)**). Different mechanical behaviors were achieved. For smaller tip angles, stiffness and strength was greatest. When considering the effect of geometry for high stiffness, strength and toughness, triangular geometry is often optimal due to its ability to uniformly distribute stress. An increment in both stiffness and strength was found as $β$ (shape factor) increased[128]. I.A. Malik *et. al* studied the pull out of jigsaw like features. Results showed that the pull-out resistance increased with the increment of interlocking angle ($\theta_0$). Due to frictional traction at the contact between the tabs, high tensile stresses are generated in the solid body (**Fig. 14(c)**)[129]. Michael M Porter *et. al* mimicked syngnathid fish structures. **Fig. 14(d).(i)** shows five models, indicated with I, II, III, IV & V. Model I is representation of a pipefish tail, Model III and V represents prehensile regions of common seahorse and pipehorse tails, respectively. Models II & IV represent intermediates models between I + III and III + V. Results show that seahorse structures display optimal performance. **Fig. 14(d).(ii)** summarizes the multifunctional performance (i.e. bending capacity, grip contact, shape restitution etc.) comparison between all 5 models[130]. The consolidated literature of bio-inspired 3D printed structures are shown in **Table 3**.

**5.4 Numerical modelling of bio-inspired structures**

Given the relative scarce variation in the basic constituent materials (essentially combinations of hard mineralized and soft tissues), the great variety of mechanical properties achieved in natural materials is mainly obtained through the complexity of structural architectures, including material arrangements, mixing, and grading over different size scales. The mechanical modelization of these complex, often hierarchical structures, requires advanced modelling tools, with the possibility of simultaneously including different length scales in simulations[131–134]. The objective of multiscale models in the mechanics of biological and bioinspired materials is thus to highlight emergent properties, from the complex hierarchical organization of multiple basic materials. This is essential, since experimental data can often be hard to rationalize, and are often related to a single size scale, i.e. can seldom provide a full understanding of the system under study.

Typically, computational models start at the nanoscale and involve ensembles of atoms or molecules. This entails the use of methods like Density Functional Theory or Molecular Dynamics.

Results of simulations at this level can be used as inputs for a variety of methods that can be used to model mechanical behaviour at the meso-scale (typically from microns to mm, or beyond). Here, the objective is to provide as simple a framework as possible, in order to treat complex architectures and accommodate hierarchy with acceptable computational costs. The methods that satisfy this requirement are mainly based on the discretization of material portions in an arrays, or networks, of basic elastic elements (linear or nonlinear) with varying constitutive properties, and statistically assigned failure thresholds, to account for defects and material heterogeneities occurring at the mesoscale. These methods include (but are not limited to) Fiber Bundle Models (FBM)[135] Lattice-Spring Models (LSM), Random Fuse Models (RFM), or Spring Block Models (SBM), to which hierarchical counterparts, or extensions, have been added, to treat hierarchical architectures.

The FBM has been widely used to simulate the behaviour of 1-D fibrous structures often occurring in Nature, such as tendons, but also in cases where materials are loaded in pure uniaxial tension. A particular size scale is modelled by using fibres arranged in series and parallel, with Weibull-distributed yield stresses, fracture strengths, or ultimate strains. When the specimen is loaded uniaxially, the weakest fibres break, and stresses are redistributed among remaining fibres according to equal (or unequal) load sharing schemes. Composite materials can be simulated using different fibre types in the bundle (**Fig. 15(a)**). A hierarchical extension to the model (Hierarchical Fibre Bundle Model – HFMB) has been introduced by using the output of multiple FBM simulations at one size scale as the input of a FBM simulation at the next scale, so that the bundle at the $i$-th hierarchical scale becomes a fibre at the next ($i$+1) hierarchical level, and the scheme can be applied recursively over all hierarchical scales constituting the structure of the analysed material. Simulations need to be repeated a sufficient number of times for statistical significance.

This type of model has allowed to model hierarchical biological materials such as spider silk[136] to derive the scaling of strength and stiffness of hierarchical structures containing defects or inhomogeneities[137] and to highlight the advantages of hierarchical structure and material mixing in terms of damage resistance[138] The same type of approach can be adopted in 2-D or 3-D versions of the FBM, namely LSM,[139] SBM[140] or RFM[141], in which the discrete elastic elements ( "springs", or "resistors") used to discretize a material volume are arranged in various types of 2D or 3D lattices. These models have proved their versatility and effectiveness in simulating effects such as plasticity, crack propagation, or statistical distributions of "avalanches" of fracture events in heterogeneous materials. Again, hierarchical extensions have been provided[142] and used in various types of problems, e.g. to simulate the failure of hierarchical nanocomposite materials[143] or the adhesive behaviour of hierarchical fibrillar structures[144] In particular, these numerical methods allow to evaluate the role of reinforcement organization, shape, aspect ratio, distribution, in avoiding direct crack path propagation and improving toughness in biological or bio-inspired composite materials[145,146]. The hierarchical SBM (**Fig. 15(b)**), on the other hand, has been used to explore the frictional properties of hierarchical multiscale structured surfaces, showing that this architecture can provide extreme tunability in the global friction coefficients and can thus potentially be exploited for industrial applications[147]. Finally, Finite-Element Methods (FEM) are one of the most widespread approaches to model the complex geometries encountered in biological and biomimetic structures[148,149]. However, the multiscale, hierarchical nature of biological materials makes the use of this numerical approach computationally very cumbersome in many cases. Additionally, while FEM approaches are usually efficient in the simulation of the emergent macroscopic elastic properties of these complex heterogeneous structures (e.g. see in [150,151], the evaluation of characteristics such as strength and toughness require reliable modelling of fracture processes, and conventional FEM approaches are not necessarily the most appropriate for this task. Nevertheless, FEM models have been used, for example, to evaluate the impact characteristics of bio-inspired armours[152,153], to study the mechanical behaviour of insect cuticle[154], or to describe the cutting mechanics of insect mandibles[155]. In the case of interpenetrating phase composites, FEM approaches have been combined with micromechanical modelling, based on homogeneization procedures that enable the calculation of effective mechanical properties, providing relatively reliable predictions for various geometries[156]. Micro-FEM based on Computed Tomography scans have been developed to allow the quantifying of bone fracture risk[157,158], and multiscale extended FEM (XFEM) approaches have been applied to predict the fracture behaviour of bone-inspired fibre-reinforced composites[159]. Furthermore, FEM approaches have been used to model bioinspired sutured materials, and their failure mechanisms occurring through pull-out of interlocked material portions[160]. Also, FEM models have been useful in describing bone tissue regeneration and growth and its interaction with biomimetic scaffolds[161].

3D printed porous structures exhibit superior mechanical strength due to complex design, higher relative density and specific material properties. There are various papers in the literature (discussed in section 3) where researchers have carried out a systematic study of the effect of geometry and relative density on mechanical properties. Specific strength and specific energy absorption are also important mechanical parameters for lightweight structures. Better mechanical strength can generally be achieved at higher relative density values. However, with the help of complex geometries, where an optimal stress distribution is possible, excellent strength values can also be achieved for lower relative density values[88,162]. In general, bio-inspired structures currently provide greater stiffness values compared to macroscopic scale based structures, mathematical model-based structures and small molecule inspired structures (**Fig. 16**). The ability to design intricate geometries with the help of mathematical algorithms and models will allow theoretical structures to compete with bio-inspired ones in the near future. On the other hand, macroscopic scale-based structures can provide relatively good mechanical properties along with stimuli-responsive behaviour, while structures inspired by small molecules can be useful in real-time applications.

## 6. Conclusions and Outlook

Merging of 3D printing and shape memory properties, along with enhanced topology can further help researchers develop new innovative materials with high mechanical performance. Origami and kirigami-inspired structures have shape memory properties that can provide controlled performance under specific conditions such as temperature and load. These architectures types can be extensively used in the fields of biotechnology, electronics and energy storage in the future. Mathematical model-inspired structures can be scaled-up by involving machine learning to help in the design of more complex topological features to enhance the capability to sustain higher mechanical loading conditions. In future, researchers can be inspired by structures such as small molecules, metal organic frameworks or aerogel foams, to design next-generation structures for enhanced mechanical performance. Mimicking biological structures has already provided new ways to enhance mechanical properties of complex bio-inspired materials, with the assistance of high-precision additive manufacturing (e.g. nacre, sea shells etc.). The bio inspired cellular or composite structures show promising mechanical robustness. Hierarchical honeycomb like structures have shown enhanced energy absorption capacities as compared to natural honeycomb structures. Similarly, bio inspired nacre-like laminated composites are found to be exceptional in resisting impulsive loading and deviating the crack propagation by controlling the ratio of soft and stiff phases. Further biomimicry studies can include inter alia muscles, DNA, sea horses, seashells, herringbone, to search for further surprising results and inspire the development of new materials and future research directions.

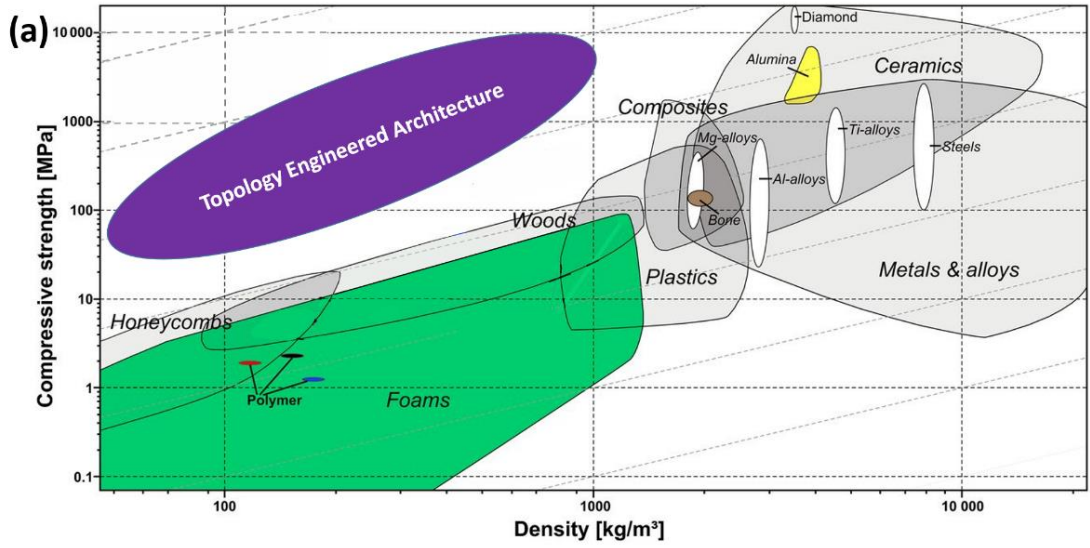

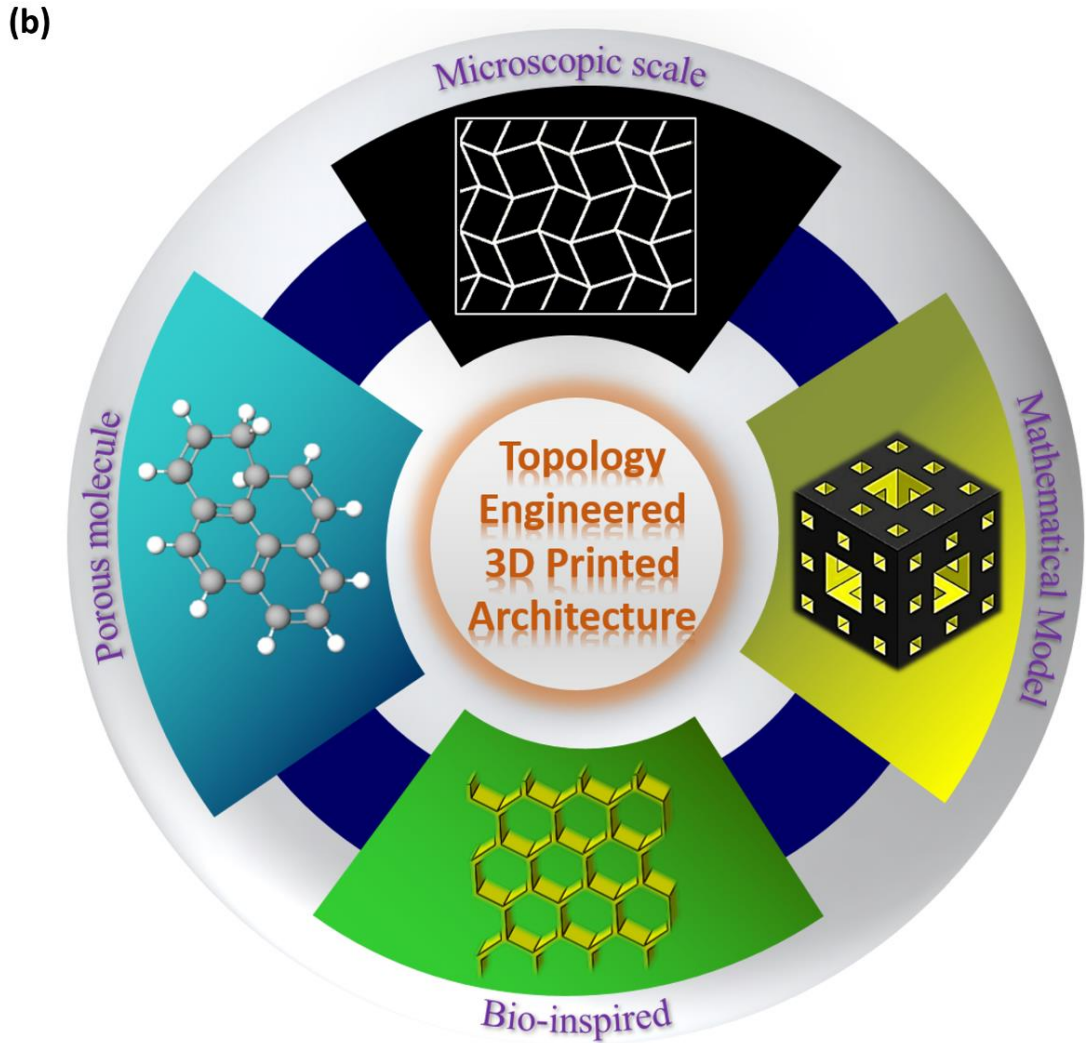

**FIGURE 1.** Overview of low-density materials along with various topology engineered structures. (a) Compressive strength–density Ashby chart for the family of materials (Reproduced with permission from [163]. Copyright © 2014, Proceedings of the National Academy of Sciences of the United States of America), (b) Classification of topology engineered architectures.

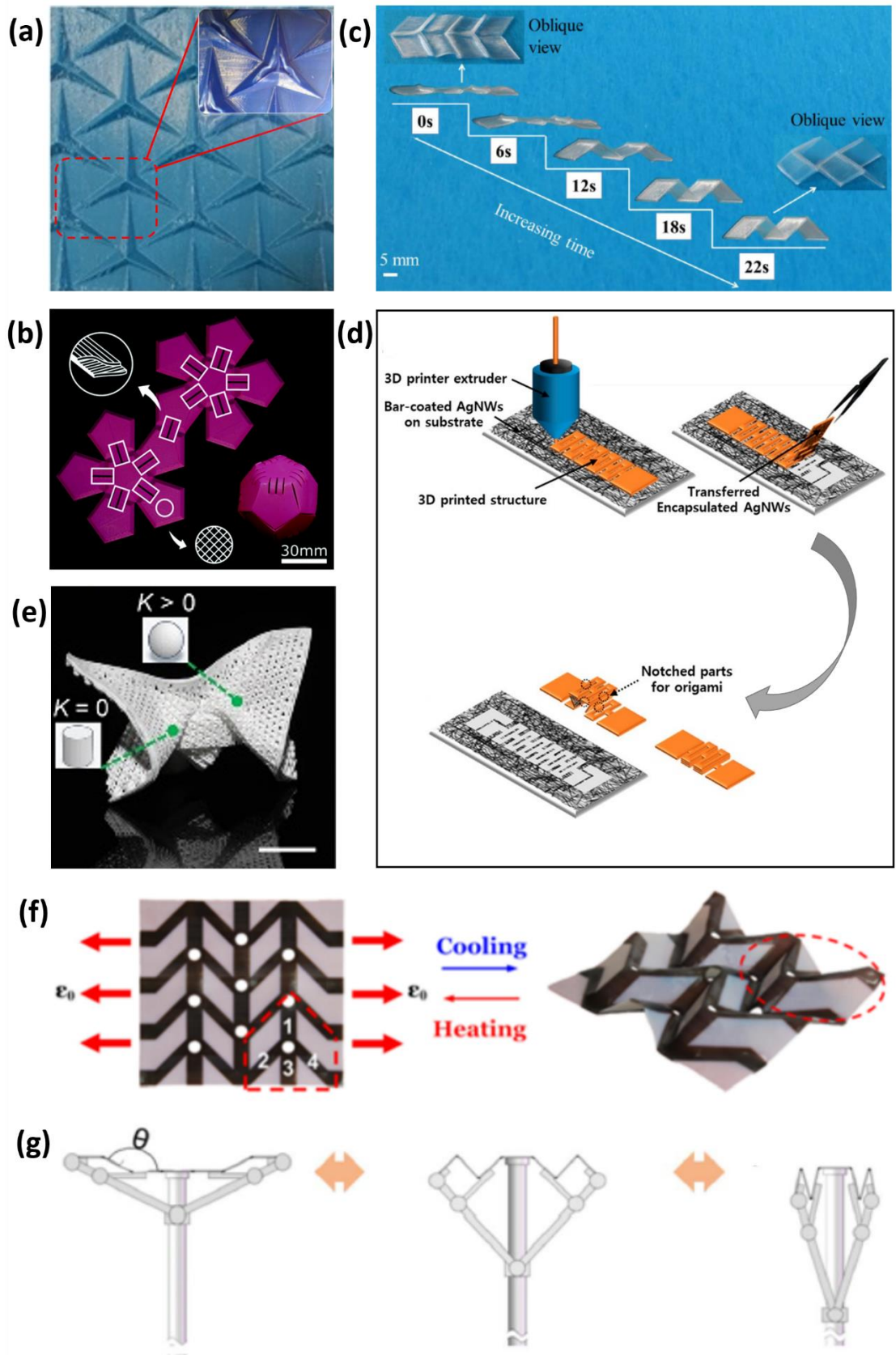

**FIGURE 2.** Origami-based structures. (a) Ron-Resch-like origami panel RR-3-15 before deformation. After plastic deformation on the tip of the star tuck (inset) (Reproduced with permission from [48]. Copyright © 2019, IOP Publishing, Ltd), (b) Self-folding of dodecanhedron into its 3D shape (Reproduced with permission from [49]. Copyright © 2017, Royal Society of Chemistry), (c) Shape recovery property of a Miura-origami tessellation structure at high temperature (90 °C) under unfolding load (Reproduced with permission from [50]. Copyright © 2018, Elsevier), (d) 3D printer-based kirigami/origami inspired structure of silver nanowires (Reproduced with permission from [51]. Copyright © 2019, American Chemical Society), (e) The inset indicates location of constraints, positive K (spherical caps) and zero K (cylinders) (Reproduced with permission from [43]. Copyright © 2018, American Association for the Advancement of Science), (f) Transformation of 2D structure into 3D Miura origami after programming[53], (g) Design of umbrella-like linkage mechanism integrated with origami substrates (Reproduced with permission from [54]. Copyright © 2019, IOP Publishing, Ltd)

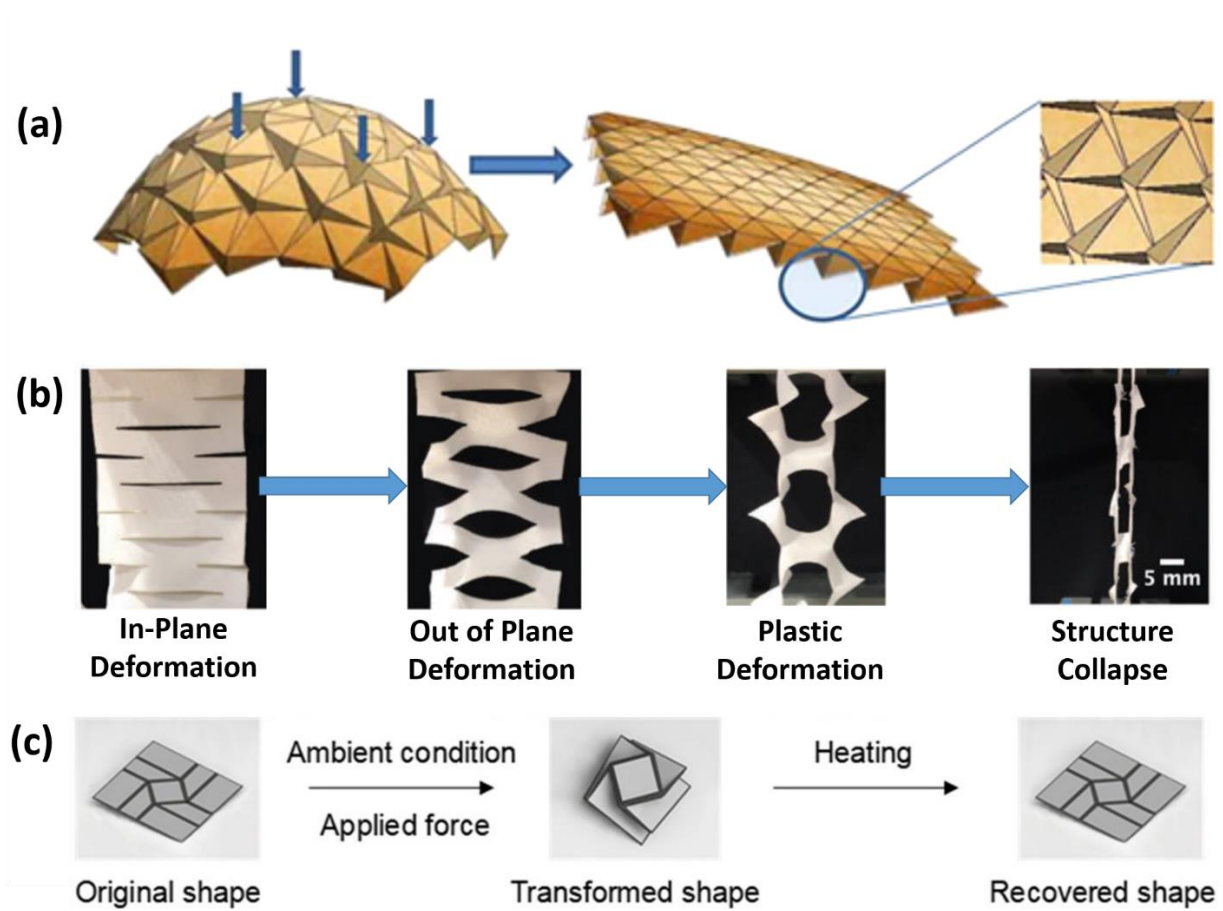

**FIGURE 3.** Kirigami-based structures. (a) Behaviour of Ron Resch dome under a compression load (Reproduced with permission from [58]. Copyright © 2019, American Society of Mechanical Engineers), (b) Optical image of detailed in-plane and out-of-plane deformation of pristine textile before 3D printing (Reproduced with permission from [60]. Copyright © 2019, John Wiley and Sons), (c) Twistable origami structure without using shape memory effect (Reproduced with permission from [45]. Copyright © 2019, American Chemical Society).

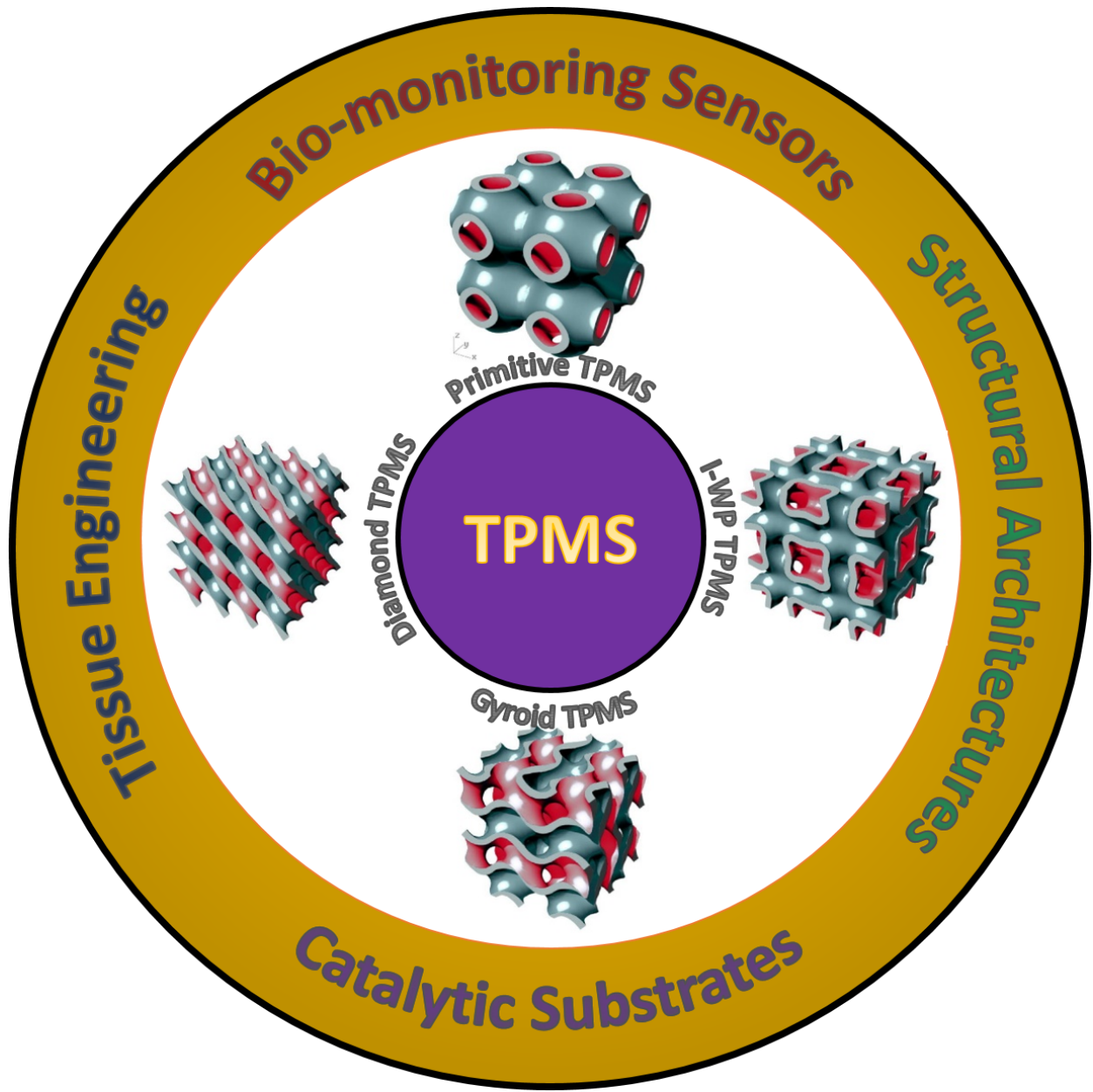

**FIGURE 4.** Classification of Triply Periodic Minimal Surface (TPMS) based structures and their applications (Reproduced with permission from [164]. Copyright © 2017, Elsevier).

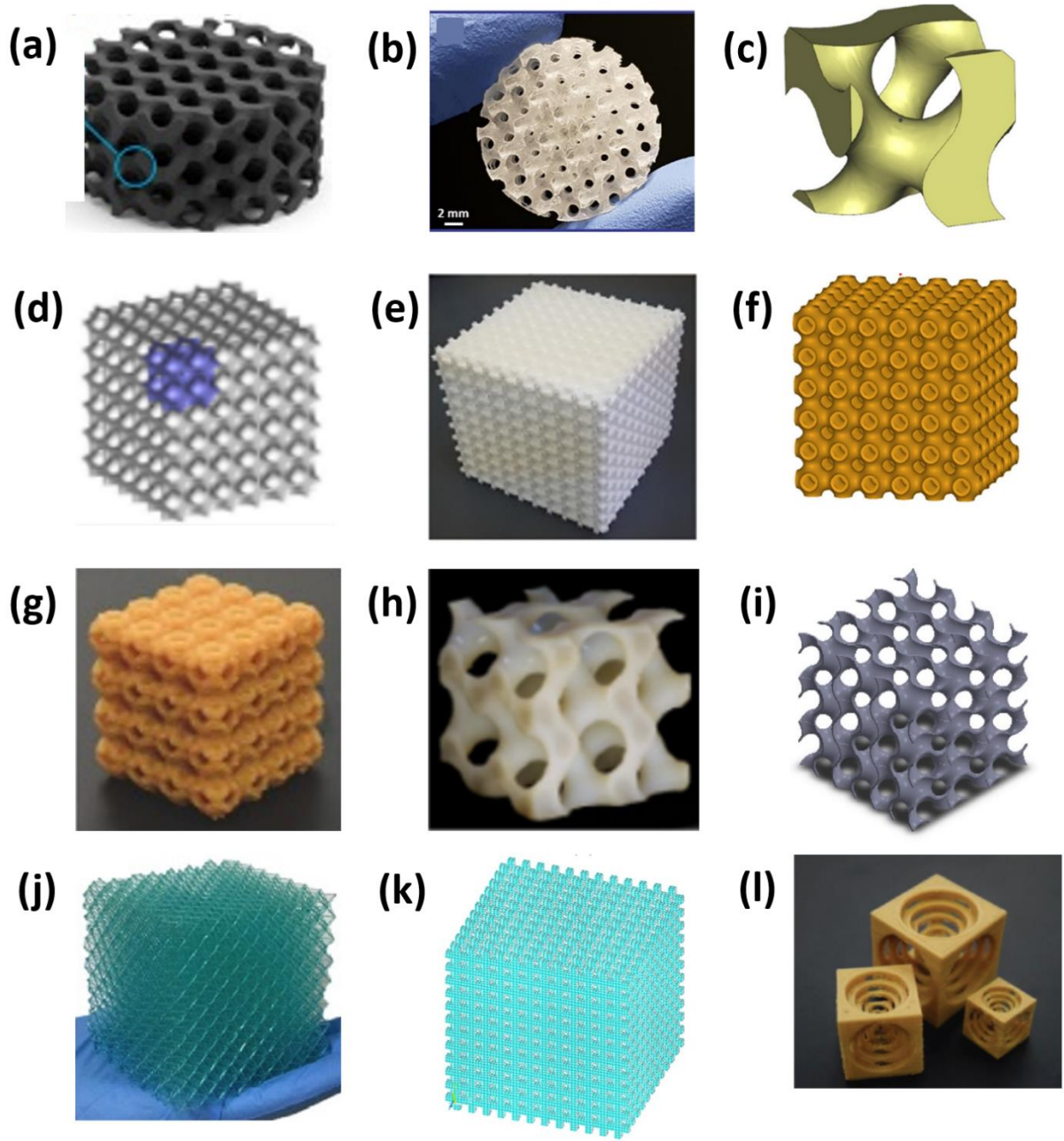

**FIGURE 5.** Mathematical model-based structures. (a) Surface embedded grapheme coating on a printed structure (Reproduced with permission from [72]. Copyright © 2020, American Chemical Society), (b) Scaffold with gradient porosity (Reproduced with permission from [73]. Copyright © 2019, Elsevier), (c) Strut-based TPMS structure (Reproduced with permission from [74]. Copyright © 2019, John Wiley and Sons), (d) Sheet-based TPMS structure (Reproduced with permission from [75]. Copyright © 2018, John Wiley and Sons), (e) Neovius cellular structure (Reproduced with permission from [76]. Copyright © 2017, Elsevier), (f) Solid lattice of primitive TPMS structure (Reproduced with permission from [77]. Copyright © 2020, Elsevier), (g) Primitive schwarzite structure (Reproduced with permission from [78]. Copyright © 2018, John

Wiley and Sons), (h) Cellular co-continuous composite (Reproduced with permission from [81]. Copyright © 2018, John Wiley and Sons), (i) cellular solid-based gyroid structure (Reproduced with permission from [69]. Copyright © 2019, Elsevier), (j) 3D fabricated octet-truss structure with 80 mm side length (Reproduced with permission from [83]. Copyright © 2018, Elsevier), (k) Cubic-iso structure (Reproduced with permission from [70]. Copyright © 2012, Elsevier), (l) Boxeption structure at multiple length scales (Reproduced with permission from [71]. Copyright © 2019, John Wiley and Sons).

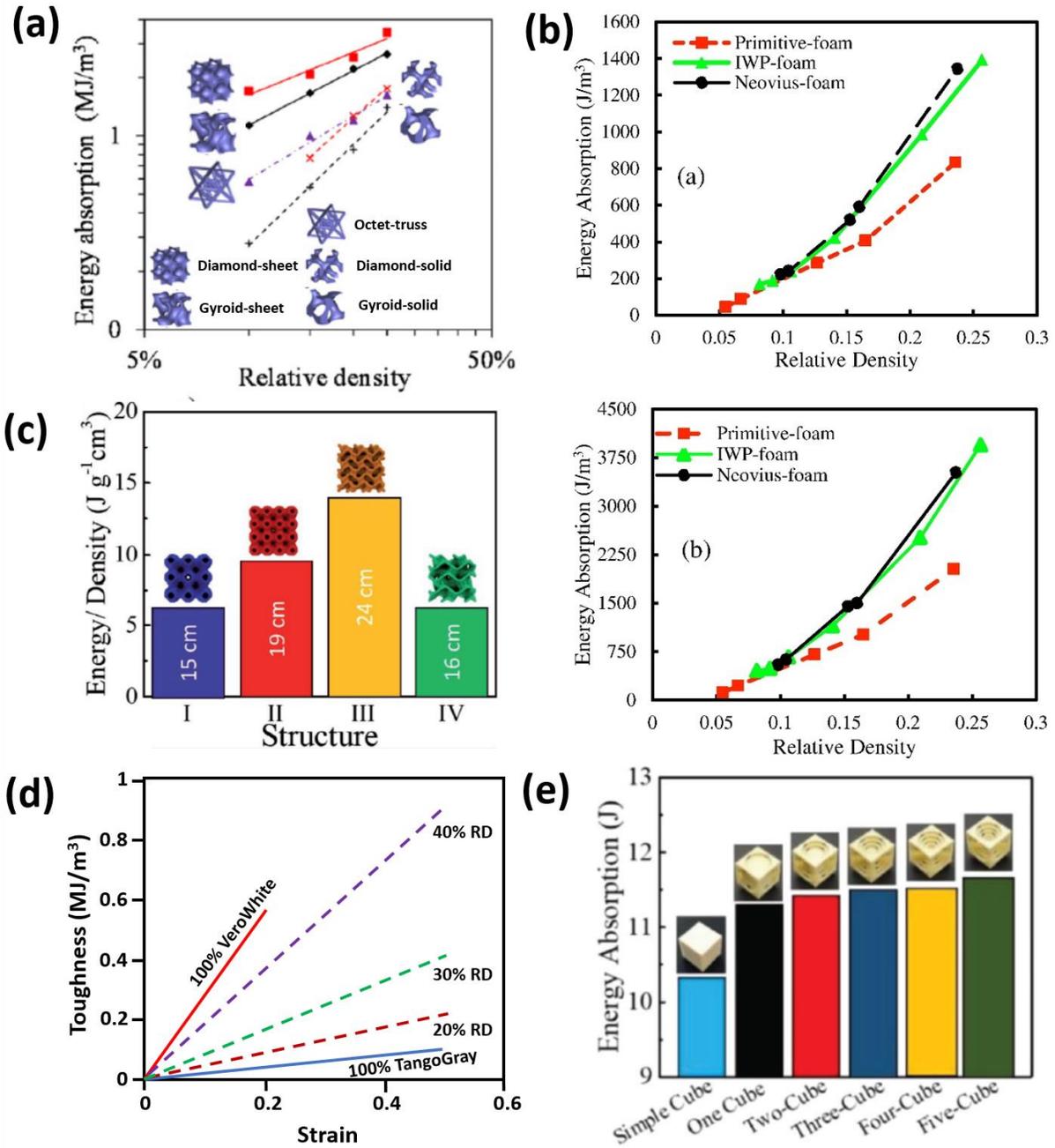

**FIGURE 6.** Energy absorption capability of mathematical model-based structures. (a) Measured energy absorption properties of Sheet-based TPMS structure as a function of the relative density (Reproduced with permission from [75]. Copyright © 2018, John Wiley and Sons), (b) Toughness of TPMS-CMs under compression of 25% strain (top) and 60% strain (bottom) (Reproduced with permission from [76]. Copyright © 2017, Elsevier), (c) Energy absorption versus density for gyroid and schwarz structures (Reproduced with permission from [78]. Copyright © 2018, John Wiley and Sons), (d) Energy absorption as a function of applied strain for a core-shell structure at high relative density (Reproduced with permission from [81]. Copyright © 2018, John Wiley and Sons), (e) Energy absorption for different boxception structures with a varying number of cubes inside the structure (Reproduced with permission from [71]. Copyright © 2019, John Wiley and Sons).

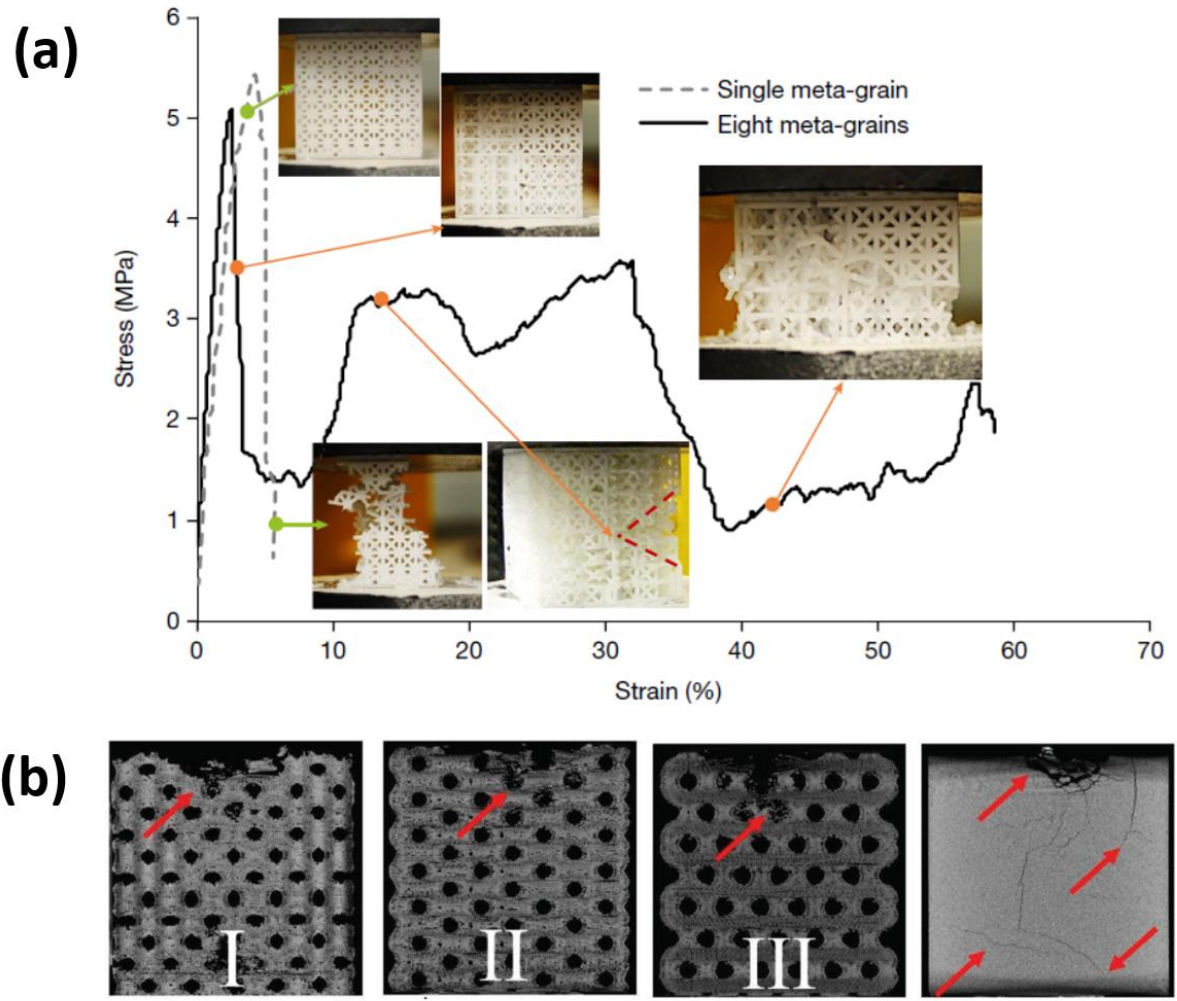

**FIGURE 7.** (a) Stress-strain curve of single and eight meta-grain structures (Reproduced with permission from [84]. Copyright © 2019, Nature), (b) CT scan snapshots of Hypervelocity impact test of tubulane structures and PLA structure (extreme right) (Reproduced with permission from [165]. Copyright © 2019, John Wiley and Sons).

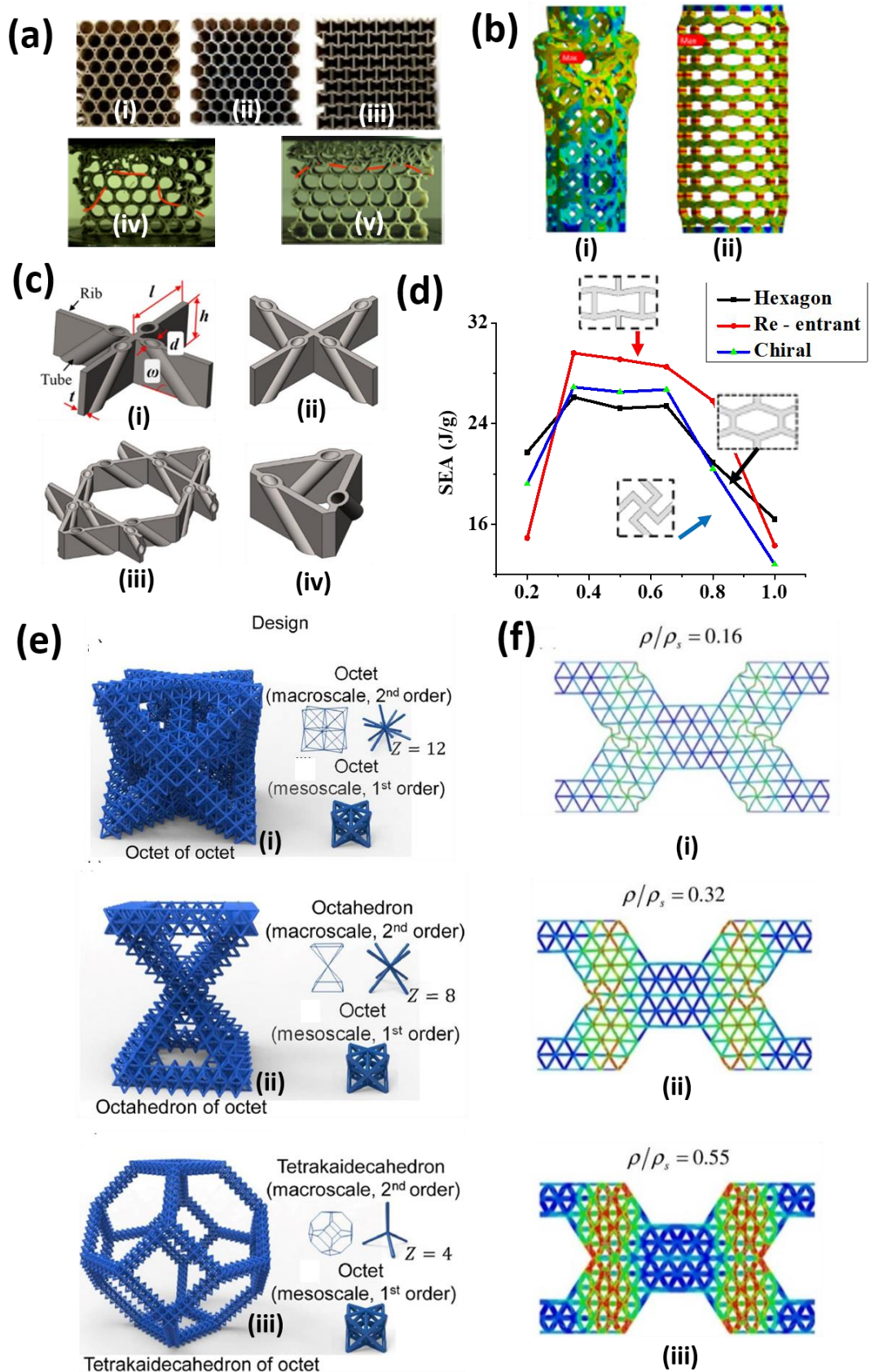

**FIGURE 8.** Bio-inspired cellular structures. (a) 3D printed honeycomb: circular, hexagonal and re-entrant (i)-(iii), Snapshots of compression behavior of circular honeycomb for a strain of ε = 0.3, (iv) PLA/KBF, (v) PLA-PCL40/KBF (Reproduced with permission from [96]. Copyright © 2019, Elsevier), (b) (i)-(ii) Numerical deformation mode of Structure I, honeycomb structure under compressive load (Reproduced with permission from [97]. Copyright © 2020, IOP Publishing, Ltd ), (c) Honeytubes structures. (i) Sq_symtube, (ii) Sq_udtube, (iii) Kag_udtube, (iv) Tri_udtube (Reproduced with permission from [98]. Copyright © 2018, Elsevier), (d) Variation of specific energy absorption (SEA) vs gradation parameter (α) (Reproduced with permission from [93]. Copyright © 2019, American Chemical Society), (e) Different hierarchical lattice materials. (i) Octet of octet (OT –OT); (ii) octahedron of octet (ON –OT). (iii) Tetrakaidecahedron of octet (TN –OT) (Reproduced with permission from [99]. Copyright © 2019, Elsevier), (f) (i)-(iii) Simulation of hierarchical honeycomb with different relative densities at a strain of $\varepsilon_y$ = -0.04. (Reproduced with permission from [91]. Copyright © 2018, Elsevier).

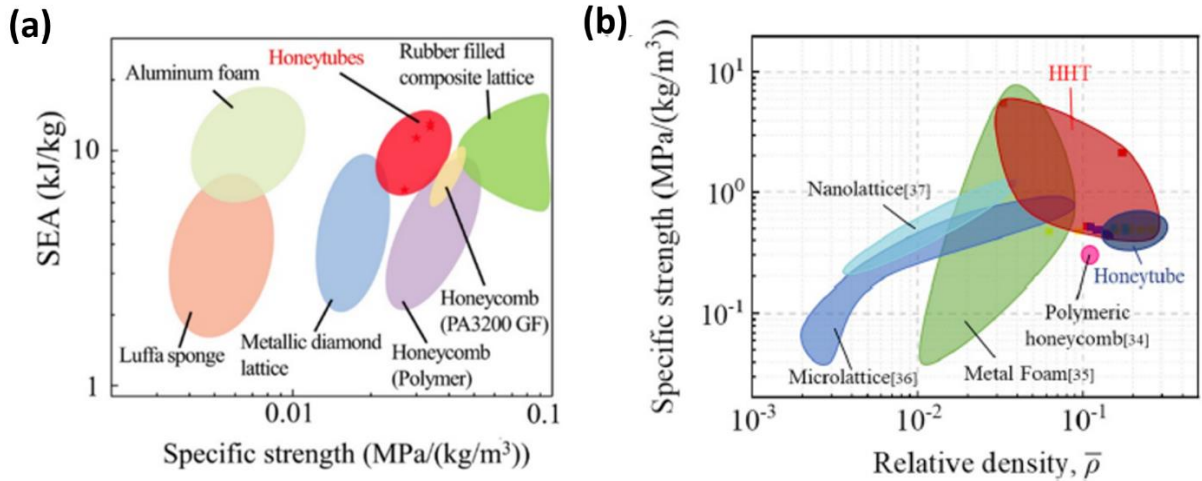

**FIGURE 9.** (a) Specific energy absorption plotted against specific strength of honeytubes compared to other lightweight cellular materials (Reproduced with permission from [98]. Copyright © 2018, Elsevier), and (b) Ashby chart for specific strength vs relative density of Hollow HoneyTubes (HHT) and its comparison with other materials (Reproduced with permission from [166]. Copyright © 2019, Elsevier).

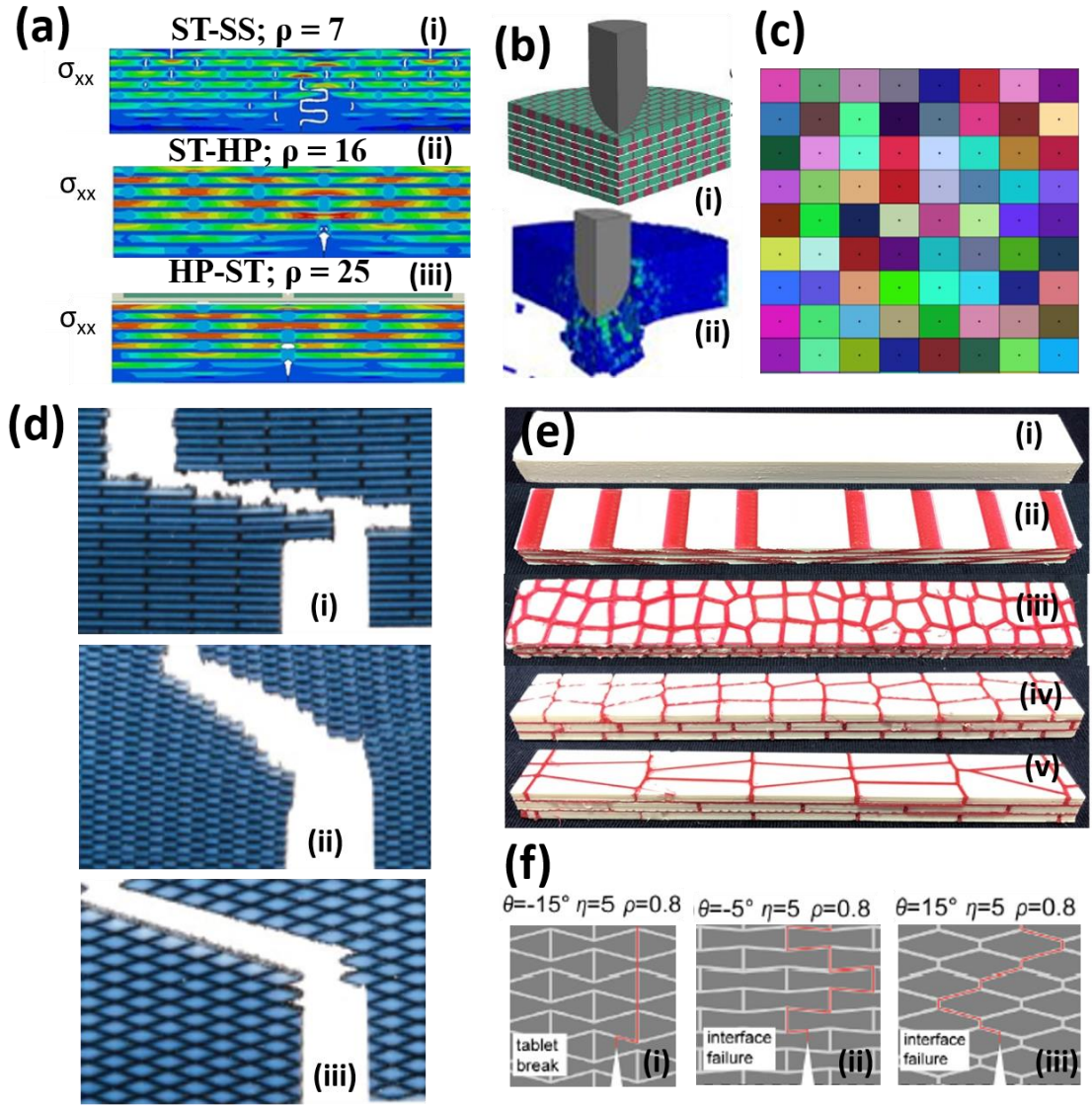

**FIGURE 10.** Bio-inspired composite structures. (a) (i)-(iii) Three staggered platelet composite specimens with different fracture responses (Reproduced with permission from [102]. Copyright © 2019, Elsevier) (b) (i) Quarter geometry of the Nacre-like design in simulations. (ii) Simulation of the impact performance of a Nacre-like sample (Reproduced with permission from [103]. Copyright © 2016, Elsevier) (c) Simple view of square Voronoi regions (Reproduced with permission from [104]. Copyright © 2017, Elsevier) (d) Schematic of investigated topologies with their fracture patterns.(i) Bone like, (ii) Bio-calcite like, (iii) rotated bone-like (Reproduced with permission from [95]. Copyright © 2013, John Wiley and Sons) (e) (i) Homogeneous PLA Specimen 1. (ii)-(iii) Nacre-like laminated composite specimens 2 and 3, without and with the use of Voronoi diagrams, respectively. (iv)-(v) Nacre-like laminated composite structure specimens 4 and 5, with doubled platelet and quadrupled platelet, respectively (Reproduced with permission from [94]. Copyright © 2019, Elsevier) (f) (i)–(iii) Schematics showing the two different failure modes: interface failure (ii) & (iii) and tablet break (i) (Reproduced with permission from [105]. Copyright © 2020, Elsevier).

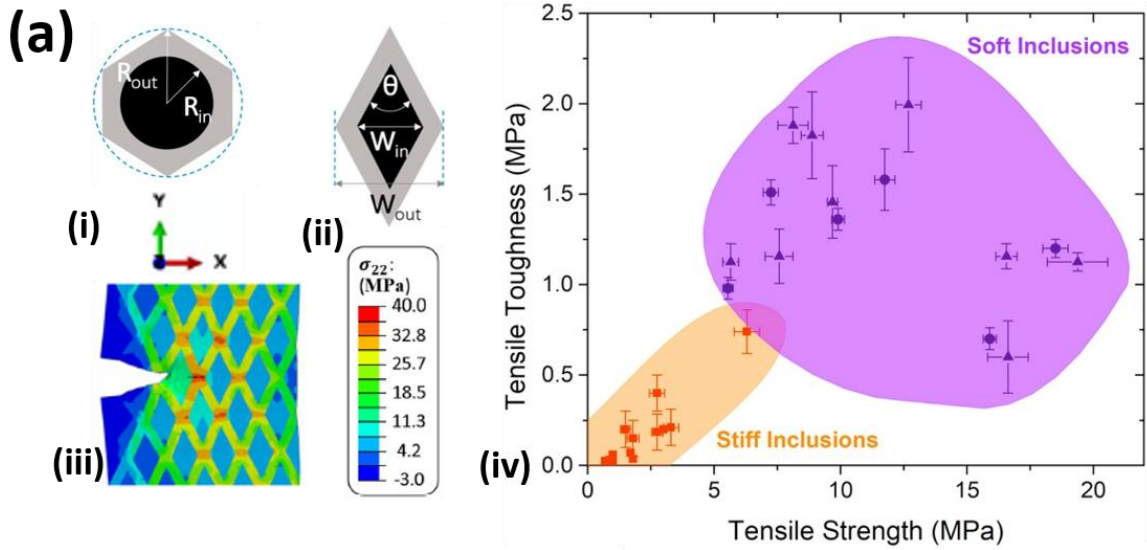
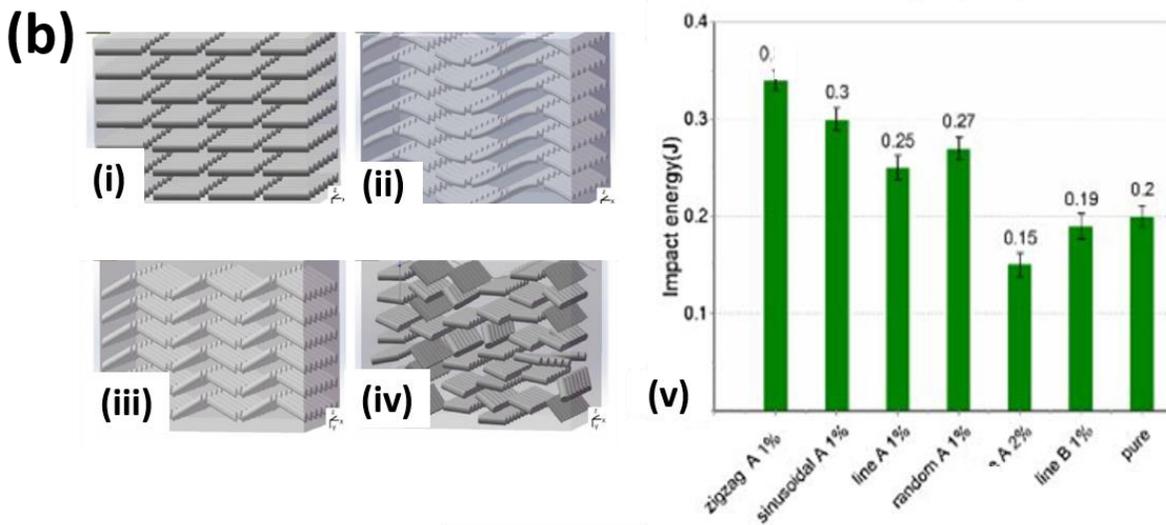
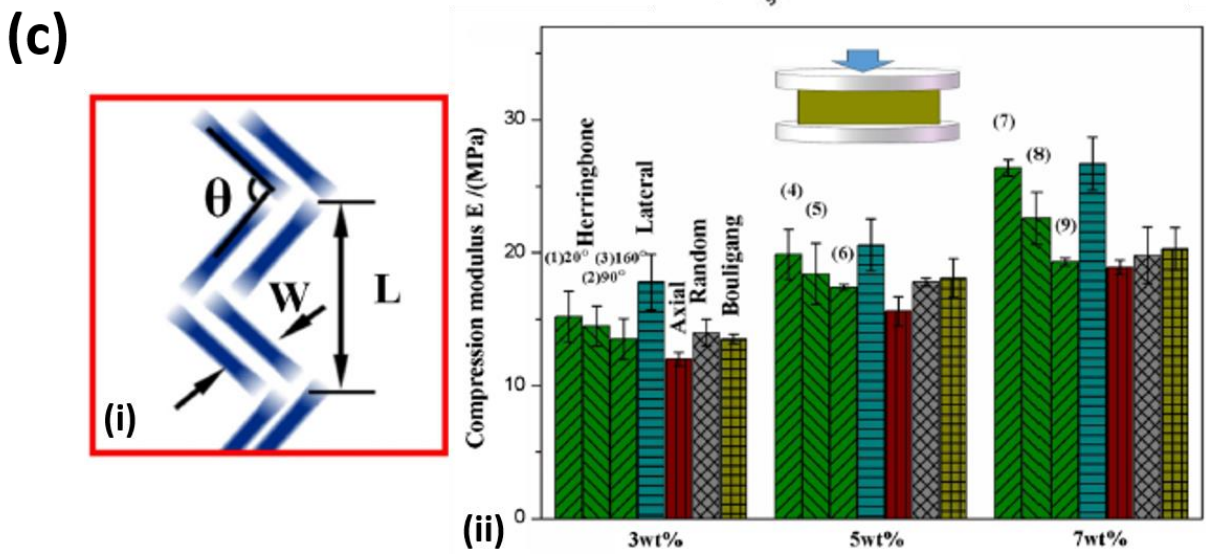

**FIGURE 11.** Bio-inspired composite structures. (a) (i)-(ii) Shape of circular and rhombic inclusions, respectively. $R_{out}$ and $R_{in}$ are the circumcircle radius of the repeat units and the radius of the inclusions, respectively. $W_{out}$ and $W_{in}$ is the short diagonal length of the repeat units and the short diagonal length of the inclusions respectively. (iii) Simulation of crack propagation in the rhombus filled composites ($\theta = \pi/3$) at 0.01/s. (iv) Chart of tensile toughness versus tensile strength of Verowhite and Tangoblack composites. Triangle points and sphere points are obtained from 20% notched samples and 40% notched samples, respectively (Reproduced with permission from [107]. Copyright © 2018, Elsevier), (b) Fibre arrangement patterns. (i) Linear, (ii) sinusoidal, (iii) zigzag, (iv) random. (v) Impact energy of the different fibre arrangements. 1% and 2% are the fibre weight fractions. "Pure" indicates matrix without fibres (Reproduced with permission from [108]. Copyright © 2019, Springer Nature) (c) (i) Schematic diagram of ''herringbone'' microarchitectures. (ii) Comparison of compressive modulus for different mass fractions or different embedded microstructures in different parts (Reproduced with permission from [109]. Copyright © 2018, Springer Nature).

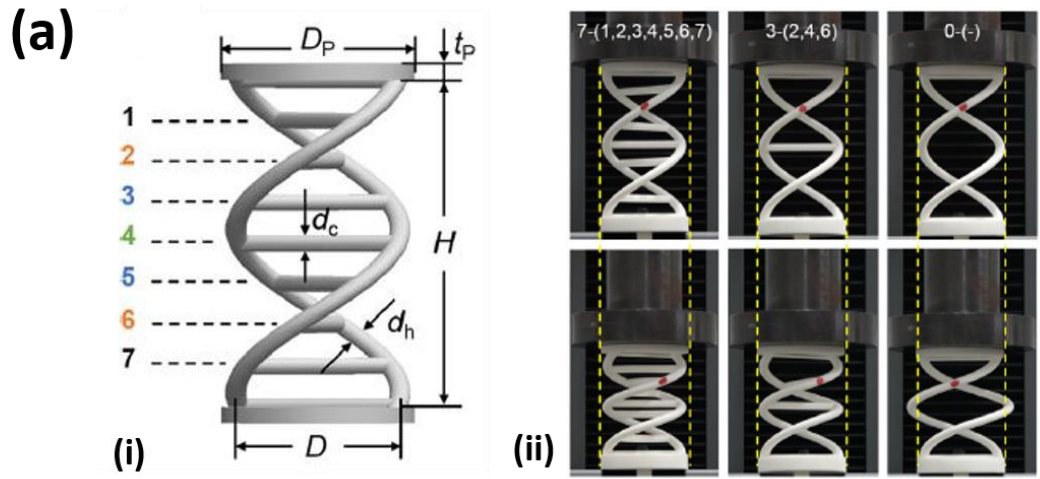
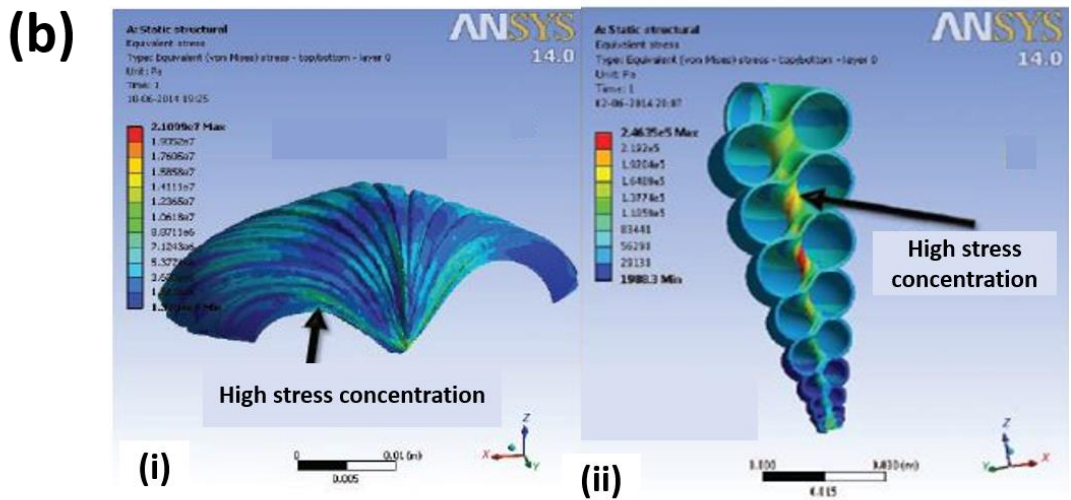
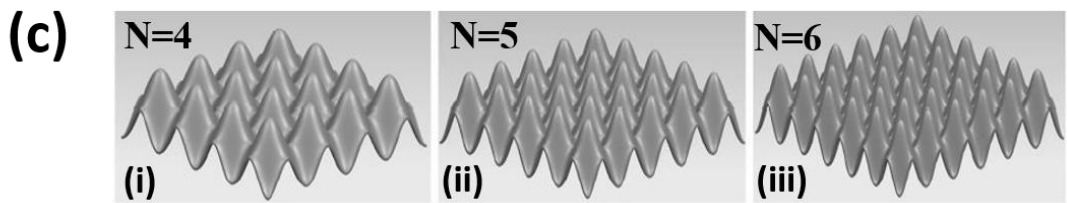
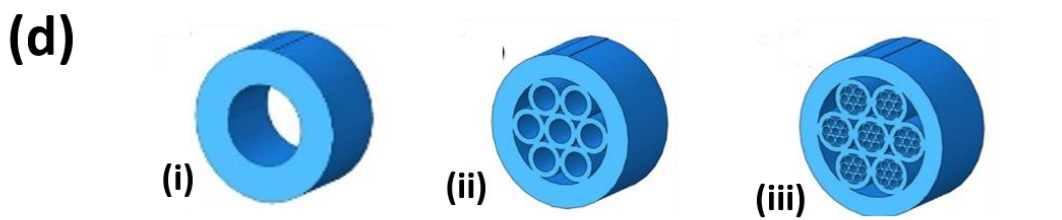
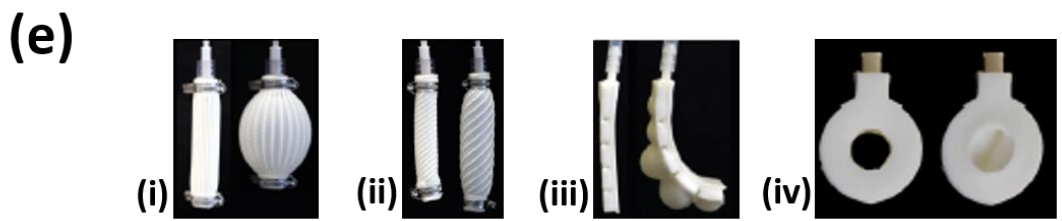

**FIGURE 12.** (a) (i) DNA-inspired helical structure. (ii) Snapshots of different DNA-inspired helical structures during compression tests (Reproduced with permission from [110]. Copyright © 2019, Elsevier), (b) Simulation. (i) Stress distribution in shell – 1, (ii) Stress distribution in shell – 2 (Reproduced with permission from [89]. Copyright © 2015, American Association for the Advancement of Science), (c) (i)-(iii) DCP (bi- directionally corrugated panel) structures with different wave numbers (N); (N=4, N=5, N=6) (Reproduced with permission from [111]. Copyright © 2020, Taylor and Francis), (d) CAD model of muscle inspired hierarchical structure: (i) 1$^{st}$ order, (ii) 2$^{nd}$ order, (iii) 3$^{rd}$ order (Reproduced with permission from [92]. Copyright © 2019, Elsevier), (e) Different modes of soft actuators (i) - (iv). (i) Contracting; (ii) Twisting; (iii) Bending; (iv) Grabbing (Reproduced with permission from [112]. Copyright © 2018, Springer Nature).

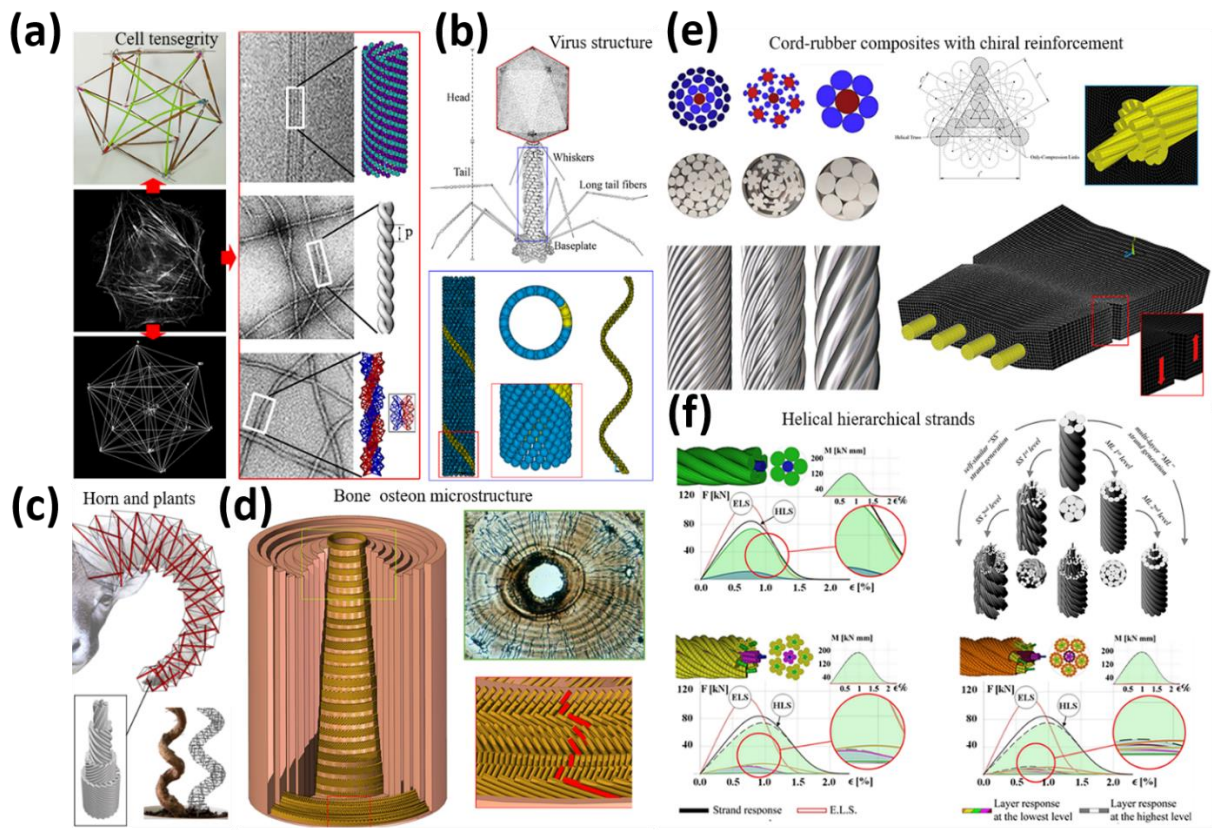

**FIGURE 13.** Examples of chirality in natural and man-made materials. (a) Tensegrity structures mimicking the cell cytoskeleton, whose constituents made of microtubules and protein filaments all exhibit a chiral structure determined by the helical geometry of polymer chains and sub-elements, which determine the mechanical response of the cell at the nano- and micro-scale and in turn for the cell adhesion, migration and duplication processes (Reproduced with permission from [113,114]. Copyright © 2019, Elsevier), (b) Helical structure of the tail of a virus, whose chiral geometry is a key mechanical feature for its functions[115]. (c) Elastic (trigonal) anisotropy and helical shape of horns and plants in which residual stress accumulated during growth, similarly to the tensegrity principle, confer optimal form and mechanical performance to the overall structure[116,117]. (d) Faithful reconstruction of the helical orientation of fibers in the lamellae of the osteon, the fundamental unit in bone tissues. (e) Rubber composites with different reinforcing cords made of (steel) helically wound wires, from which the strength at the interface in the operational conditions depends (Reproduced with permission from [124]. Copyright © 2021, Elsevier), (f) Hierarchical chiral strands with various microstructural helical configurations to obtain the so-called Equal Load Sharing (ELS) effect that allows the structure to recover its optimal mechanical response by homogeneously distributing the stress as one or more wires break.

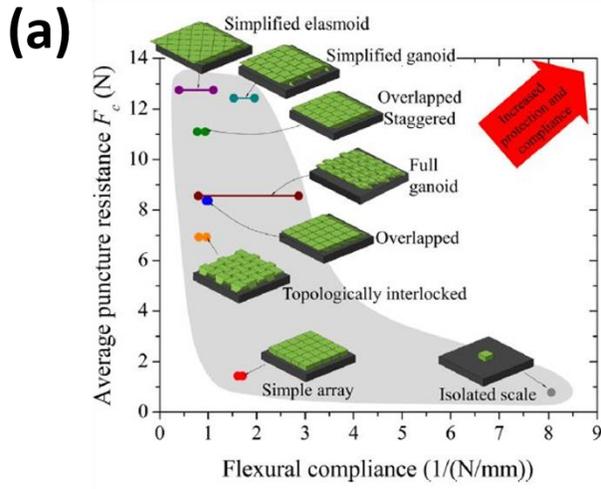
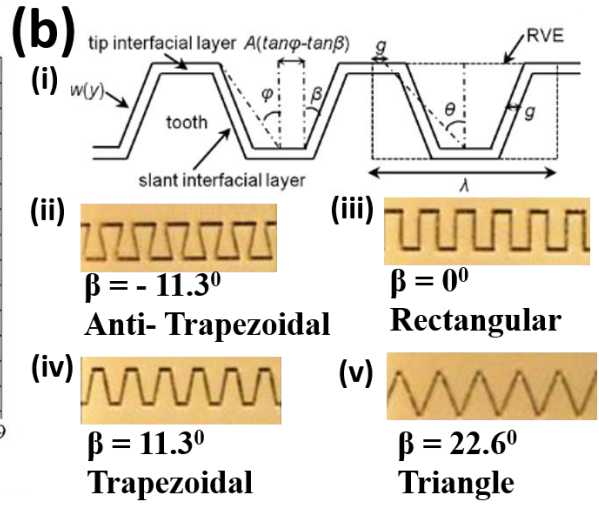
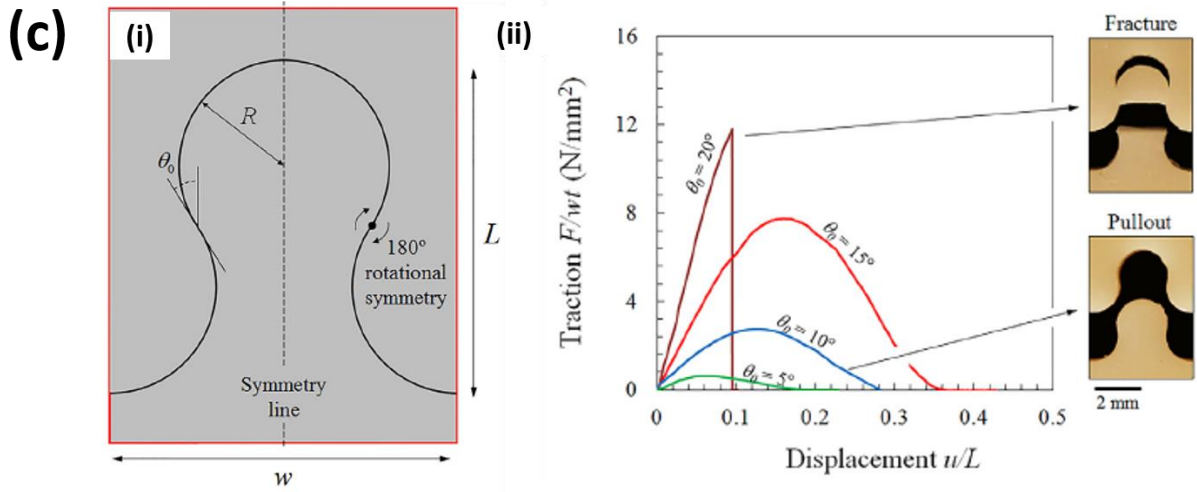
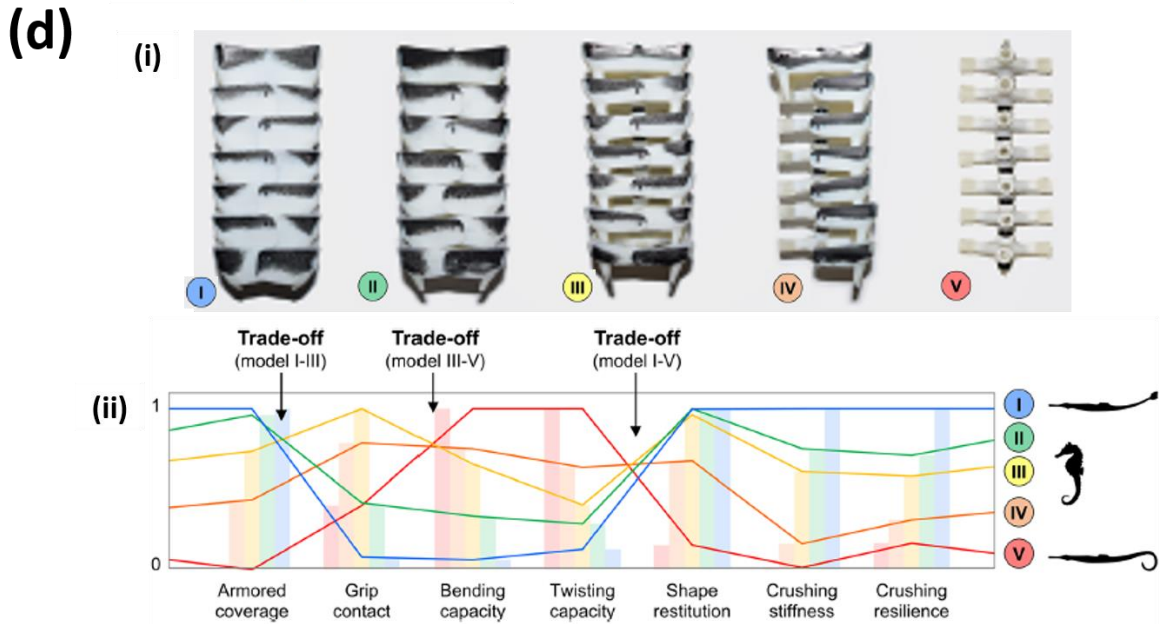

**FIGURE 14.** (a) Ashby plot showing the combination of flexural compliance vs puncture resistance for different designs (Reproduced with permission from [127]. Copyright © 2017, Elsevier), (b) (i) Schematic diagram of suture interfaces with their geometrical parameters. (ii)-(iv) Optical images of suture interface prototypes (Reproduced with permission from [128]. Copyright © 2014, Elsevier), (c) (i) Geometry of unit cell of jigsaw locking features, (ii) Representative pull-out curves with different interlocking angles, showing different failure modes: Tab pull-out and fracture (Reproduced with permission from [129]. Copyright © 2017, Elsevier), (d) (i) Image of ink-stamped contact areas of the five models. (ii) Comparison of multifunctional performance of models I–V (Reproduced with permission from [130]. Copyright © 2017, IOP Publishing, Ltd).

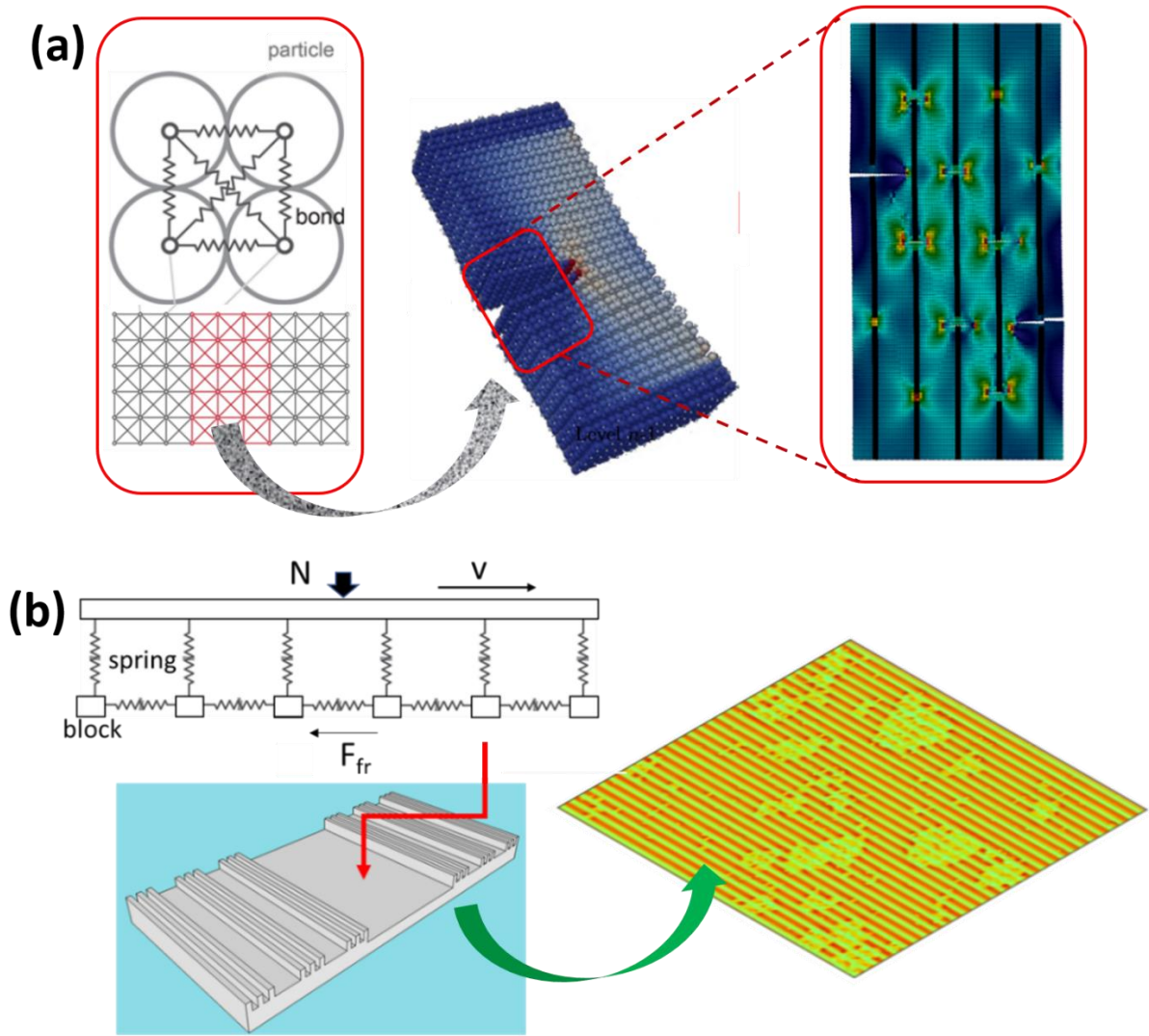

**FIGURE 15.** a) Hierarchical lattice spring to simulate the fracture of fibre-reinforced composite materials, highlighting the effect of load concentrations and crack stopping effects; b) Spring block

model to simulate the friction of hierarchically rough surfaces, highlighting the propagation of surface detachment fronts.

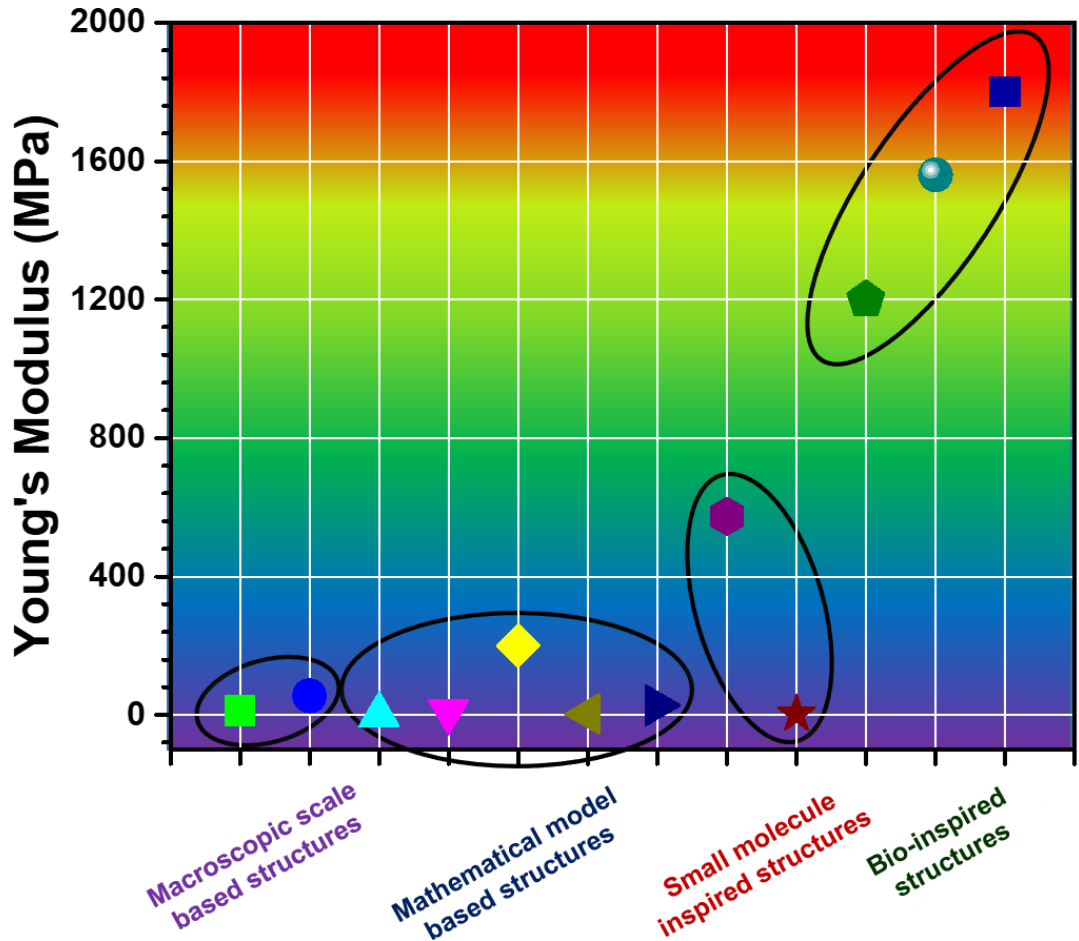

**FIGURE 16.** Young's modulus of 3D printed structures for topology enhanced strengthening[48,59,110,167,71,73,78,81,82,86,91,93].

**TABLE 1.** Origami/kirigami-inspired 3D printed structures.

| Inspiration | Materials | Technique | Application | Ref. |
|---|---|---|---|---|
| Ron Resch-like origami | Polylactic acid (PLA) | FDM | Load and energy damper | [48] |
| Origami | PLA | FDM | Biotechnology and electronics | [49] |
| Origami | PLA based shape memory polymer | FDM | Actuators and reconfigurable devices | [50] |
| Origami | Thermoplastic polyurethane (TPU) | FDM | Highly flexible future energy conversion | [51] |
| Origami | Silicone adhesive [DOWSIL™ SE 1700] | Direct ink writing | - | [52] |
| Origami | Poly(dimethylsiloxane) (PDMS) and $ZrO_2$ nanoparticles | Inkjet printing | Autonomous morphing structures, aerospace propulsion components, space exploration, electronic devices, and high-temperature microelectromechanical systems | [43] |
| Origami | Tangoblack and Verowhite | Multi-material inkjet printing | Load-bearing application | [53] |
| Origami | Tangoblack and Verowhite | Multi-material inkjet printing | Consecutive frequency-reconfigurable antennas | [54] |
| Origami | Tangoblack and Verowhite | Multi-material inkjet printing | Self-assembly | [55] |
| Origami | n-type CNT ink and Polyvinylpyrrolidone | Multi-material inkjet printing | Nondestructive inspection | [56] |
| Square-twist origami | VeroWhitePlusPlus and TangoBlackPlus | Multi-material polyjet printing | Metamaterials and robotics | [34] |
| Origami | Aliphatic urethane diacrylate/ glycidylmethacrylate/isodecyl acrylate based resin | Digital light projector 3D printer | Soft robots, stretchable electronics, and mechanical metamaterials | [44] |
| Origami | PDMS | Mask-image-projection-based stereolithography process | Biomedical and electronics | [57] |
| Ron Resch-like origami | Nylon | Selective laser sintering (SLS) | Energy absorption application | [58] |
| Origami | Stainless steel powders (316L) | Selective laser melting | High-temperature low electromagnetic reflectivity | [59] |
| Kirigami | TPU | FDM | Wearable devices | [41] |
| Kirigami | Reactive silver inks | Multi-material Inkjet printing | Wearable textile electronics | [60] |
| Kirigami | Silver nanoparticles | Multi-material Inkjet printing | Strain sensor | [61] |
| Kirigami | Polyvinylidene fluoride (PVDF) and Multi-walled carbon nanotubes powder VeroWhitePlus | Multi-material Inkjet printing | Electrodes for flexible and deformable Li-Ion batteries | [62] |
| Kirigami | VeroWhitePlus TangoBlackPlus | Multimaterial Polyjet printing | Self-releasing, self-sinking, light switch, mechanical energy storage and mechanical actuators | [45] |

**TABLE 2.** Mathematical model, microstructure and small molecule-inspired 3D printed structures.

| Inspiration | Materials | Technique | Application | Ref. |
|---|---|---|---|---|
| Triply periodic minimal surfaces (TPMS) | Acrylonitrile butadiene styrene (ABS) | FDM | Sensors for wearable biomonitoring | [72] |
| TPMS | PLA | FDM | Tissue regeneration | [73] |
| TPMS | Acrylic resin | 3D printing | Catalytic substrate | [74] |
| TPMS | Polymeric material | Direct laser writing | - | [75] |
| TPMS | PA 2200 | Selective laser sintering (SLS) | Automotive and aerospace industry | [76] |
| Schwarz Primitive TPMS | PA2200 | Selective laser sintering (SLS) | Engineering structural application | [77] |
| Primitive and gyroid Schwarzite | PLA | Multi-material 3D printing | Mechanical dampers and impact resistance in automotive, aerospace, and defense application | [78] |
| Schwartz diamond | Titanium | Laser powder bed fusion | Bone implant | [79] |
| Schwartz diamond | Titanium | Selective laser melting | Bone implant | [80] |
| Periodic Gyroidal | VeroWhite and TangoGray | Polyjet printing | Mechanical energy absorber | [81] |
| Gyroid | Polyamide 12 | Selective laser sintering (SLS) | Structural architecture | [69] |
| Octet-truss, gyroid, cubo-octahedron, and Kelvin | Graphene oxide, Photocurable acrylates and photoinitiator | Spatial light modulator | Energy storage and conversion, separations, and catalysis | [82] |

| Octet-truss | UV-curable photopolymer | Stereolithography | Energy absorption | [83] |
| --- | --- | --- | --- | --- |
| Cubic, cubic-iso and hexagonal | VeroWhitee- FullCure 830 | Inkjet printing | Bone tissue engineering | [70] |
| Boxception | PLA and polyvinyl alcohol | Multi-material printing | Sandwiched structures | [71] |
| Crystal microstructure | PLA and thermoplastic co-polyester | FDM | Damage-tolerant architectures | [84] |
| Cross-linked carbon nanotube | PLA | Multi-material printing | Hypervelocity impact resistant structures | [165] |
| Agglomerate | VeroWhitePlus™ and DM 9895 | Polyjet printing | - | [86] |
| Zeolite | PLA | FDM | Load bearing application | [87] |
| Body centered cube | Carbon nanotube/PA12 | SLS | Cushion, impact protection, explosion-proof, protective packaging | [88] |

**TABLE 3.** Bio-inspired 3D printed structures.

| Inspiration | Materials | Technique | Application | Ref. |
| --- | --- | --- | --- | --- |
| Honeycomb | PLA,PCL(Polycaprolactone),KBF(silane-treated basalt fiber) | FDM | Load-bearing and energy absorption | [96] |
| Glass sponge | - | - | Lightweight structures for aerospace, vehicle and ships | [97] |
| Honeycomb (honeytubes) | PA3200 GF, a whitish, glass-filled polyamide 12 powder | SLS | Crash protection | [98] |
| Honeycomb | VeroWhite and TangoPlus | Polyjet printing | Heel of shoe | [93] |

| Hierarchical structures | Nylon 12 | SLS | - | [99] |
|---|---|---|---|---|
| Honeycomb | VeroWhite | - | Low-velocity impact and shock wave resistance structures | [91] |
| Staggered platelet | Verowhite & DM9895 | Polyjet printing | - | [168] |
| Nacre | Veromagenta and Tangoblackplus | - | - | [103] |
| Nacre | ABS, PLA and TPU | FDM | Shockwave energy absorber | [169] |
| Nacre and bone | VeroWhitePlus and TangoBlackPlus | Dual material jetting technology | - | [95] |
| Nacre | VeroWhite and TangoPlus | Polyjet printing | - | [105] |
| Honeycomb | VeroWhitePlus and TangoBlackPlus | - | Load transfer | [107] |
| Herringbone-modified helicoidal architecture of mantis shrimp | Photosensitive resin and nickel coated carbon fibers | Magnetic printing | Artificial materials | [170] |
| DNA | PA2200 | - | Bio-inspired mechanical metamaterials and impact energy absorbers | [110] |
| Sea shells | PLA | FDM | - | [171] |
| Telson of mantis shrimp | AlSi$_{10}$Mg powder | SLM | - | [172] |
| Muscle | TPU | FDM | Energy absorption system | [92] |

| Muscular Hydrodstat | EcoFlex 00–30 A and DragonSkin 30 A, Sylgard 184, Pentaerythritol Tetrakis, 1-Hydroxycyclohexyl phenyl ketone | Direct ink writing | Robotic soft actuators | [112] |
|---|---|---|---|---|
| Fish scale | ABS photopolymer | Direct light projector | Puncture resistance protective system | [127] |
| Suture interface | VeroWhite and TangoPlus | Polyjet printing | - | [128] |
| Bone, teeth and mollusc shell | ABS | Digital light processing | Damp and shock absorber | [129] |
| Seahorse and piphorse | VeroWhite and TangoPlus | Polyjet printing | Bomb disposal robots | [130] |